\documentclass[aps,preprint,groupedaddress,graphicx]{revtex4-1}
\pdfoutput=1 

\usepackage{amsmath,amssymb,amstext,bm}
\usepackage[pdftex]{graphicx} 
\usepackage{siunitx}
\DeclareSIUnit\node{node}

\usepackage{subcaption}
\graphicspath{{images/}}

\usepackage{hyperref}
\usepackage{breakurl}

\begin{document}

\title{Field-driven dynamics of nematic microcapillaries}

\author{Pouya Khayyatzadeh}
\author{Fred Fu}
\author{Nasser Mohieddin Abukhdeir}
\email{nmabukhdeir@uwaterloo.ca}
\homepage{http://chemeng.uwaterloo.ca/abukhdeir/}
\altaffiliation{Waterloo Institute for Nanotechnology, University of Waterloo, Waterloo, Ontario, Canada}\
\affiliation{Department of Chemical Engineering, University of Waterloo, Waterloo, Ontario, Canada}

\date{\today}

\begin{abstract}
Polymer-dispersed liquid crystal (PDLC) composites have long been a focus of study for their unique electro-optical properties which have resulted in various applications such as switchable (transparent/translucent) windows.
These composites are manufactured using desirable ``bottom-up'' techniques, such as phase separation of a liquid crystal/polymer mixture, which enable production of PDLC films at very large scales.
LC domains within PDLCs are typically spheroidal, as opposed to rectangular for an LCD panel, and thus exhibit substantially different behaviour in the presence of an external field.
The fundamental difference between spheroidal and rectangular nematic domains is that the former results in the presence of nanoscale orientational defects in LC order while the latter does not.
Progress in the development and optimization of PDLC electro-optical properties has progressed at a relatively slow pace due to this increased complexity.
In this work, continuum simulations are performed in order to capture the complex formation and electric field-driven switching dynamics of approximations of PDLC domains.
Using a simplified elliptic cylinder (microcapillary) geometry as an approximation of spheroidal PDLC domains, the effects of geometry (aspect ratio), surface anchoring, and external field strength are studied through the use of the Landau--de Gennes model of the nematic LC phase.
\end{abstract}


\maketitle

%%%%%%%%%%%%%%%%%%%%%%%%%%%%%%%%%%%%%%%%%%%%%%%%%%%%%%%%%%%%%%%%%%%%%%%%%%%%%%%%%%%%%%%
\section{Introduction}

Traditional liquid crystal display (LCD) technology is principally based upon manipulating optical properties of nematic liquid crystal (LC) thin films through the application of spatially localized electric fields.
LCDs utilize the combination of a rectangular thin film geometry, precisely controlled surface anchoring conditions, and the application of an electric field in order to manipulate a \emph{defect-free} LC texture into a desired state \cite{Drzaic1995,Bronnikov2013}.
A key aspect of traditional LCD technology is that conditions are engineered such that they do not impose topological constraints which result in the formation of orientational defects.
Alternatively, an increasing number of different LC mechanisms and phases have been discovered which leverage the presence of orientational defects \cite{Serra2011}.
 
Polymer-dispersed liquid crystal (PDLC) films are one of the most studied LC-based materials which involve LC dynamics with orientational defects present.
One of the main applications of these films is as switchable ``privacy glass'', where an electro-optical property of the film is used such that an applied field drives the film to a transparent state and, upon release of the field, the film returns to a translucent state \cite{Drzaic1995}.
PDLC composites are formed through ``bottom-up'' manufacturing processes, mainly through photopolymerization-induced phase separation \cite{Drzaic1995}.
For low volume fractions of the LC component, the resulting composite morphology involves a spheroidally-confined dispersed LC guest phase in a polymer matrix phase.
This spheroidal confinement imposes topological constraints on the dispersed LC domains which require the formation of orientational defects.
Experimental observation of these PDLC films in the presence of external fields has shown a rich and complex range of electro-optic behavior \cite{Serra2011} depending on PDLC composition, chemistry, structure, polymer/LC anchoring conditions, and external field strength.

Experimental studies have indicated that non-spherical droplet shape, specifically anisometric shape, is a key factor in PDLC film relaxation time following release of the external field \cite{Drzaic1988}.
However, direct experimental observation of LC internal structure and dynamics for PDLC composites is challenging due to the length ($\si{\nano\metre}-\si{\micro\metre}$) and time ($\si{\nano\second}-\si{\micro\second}$) scales involved in LC dynamics.
Simulation studies \cite{Chiccoli1990,Berggren1992,Chiccoli1996,Smondyrev1999,Gartland1991,Sonnet1995,DeLuca2007,Ding1995,Li1999,Kanke2013,Chan1997b,Chan1999,Bharadwaj2000,Chan2001,Chan2001a,Rudyak2013,Yan2002,Sharma2003}, however, have shed light on a far more rich landscape of internal structure than what is observable through experimentation.
Over the past two decades, simulation-based analysis has been used with increasing success, mainly focusing on cylindrical and spherical domains.
Lattice-based simulations \cite{Chiccoli1990,Berggren1992,Chiccoli1996,Smondyrev1999} are able resolve sub-micron domains and have mainly been applied to study the effects of sub-micron cylindrical confinement where geometry and anchoring affects the stability of the nematic phase.
However, continuum simulations \cite{Ding1995,Li1999,Kanke2013,Chan1997b,Chan1999,Bharadwaj2000,Chan2001,Chan2001a,Rudyak2013,Yan2002,Sharma2003,Abukhdeir2009,Soule2009a} have been able to overcome the length and timescales required to simultaneously capture defect dynamics (nanoscale) and domain shape ($\ge \si{\micro\metre}$).

Continuum simulations of confined LC domains \cite{Ding1995,Li1999,Kanke2013,Chan1997b,Chan1999,Bharadwaj2000,Chan2001,Chan2001a,Rudyak2013,Yan2002,Sharma2003,Abukhdeir2009,Soule2009a} have been conducted using either Frank--Oseen vector theory \cite{Kralj1995} or Landau--de Gennes tensor theory \cite{Gartland1991,Sonnet1995,DeLuca2007}.
While many of these past studies have focused on circular/spheroidal \cite{Ding1995,Li1999,Kanke2013} and elliptic/ellipsoidal \cite{Chan1997b,Chan1999,Bharadwaj2000,Chan2001,Chan2001a,Rudyak2013} confined LC domains, they have relied on Frank--Oseen theory which cannot capture orientational defects and phase transition.
More recently, simulations of nematic LC confinement have been performed using the high-descriptive Landau--de Gennes tensor theory \cite{Yan2002,Sharma2003}, but these studies were limited to cylindrical domains and in the absence of an external field.

In this context, the overall objective of this study is to predict both the formation and electric field-driven dynamics of nematic elliptic cylinder domains.
While this geometry is a poor approximation of spheroidal and ellipsoidal domains observed in PDLCs \cite{Drzaic1988}, it has direct relevance to the study of nematic-filled capillaries which are of interest for fiber optics-based devices \cite{Warenghem1998}.
Furthermore, even though two-dimensional elliptic domains are not topologically equivalent to three-dimensional ellipsoidal domains, this ``coarse'' geometric simplification has been used in almost all past simulation-based work in the area except for in Ref. \cite{Rudyak2013}.
The Landau--de Gennes tensor model for the nematic phase is used in order to capture the presence of orientational defects, experimentally-relevant anchoring conditions, and phase transition.
The specific objectives are to study the effects of geometry (aspect ratio) and anchoring conditions on:
\begin{itemize}
    \item the formation dynamics and equilibrium textures of nematic elliptic cylinder domains. This is similar to the state of the PDLC after quenching the phase-separated film.
    \item the external field-driven dynamics of nematic elliptic cylinder domains. This similar to the state of the PDLC after application of an electric field to induce the transparent state.
    \item the relaxation or ``restoration'' dynamics of nematic elliptic cylinder domains following release of the external field. This is similar to state of the PDLC after release of the electric field to return to the translucent state.
\end{itemize}
This study is performed with a few important assumptions; the hydrodynamic effects and thermal fluctuations of nematic order are neglected.
Additionally, heterogeneous nucleation is assumed to be the dominant mechanism for formation of the domain from the disordered/isotropic phase, which is based on recent experimental observations \cite{Aya2011}.

The paper is organized as follows: first the theoretical and numerical bases for the simulation method are presented, next the nematic domain visualization and quantification methods are described, then the results are presented and discussed for formation and switching dynamics, and conclusions made.

%%%%%%%%%%%%%%%%%%%%%%%%%%%%%%%%%%%%%%%%%%%%%%%%%%%%%%%%%%%%%%%%%%%%%%%%%%%%%%%%%%%%%%%
\section{Background}

In this section the theoretical model and simulation methods are briefly described.
The theoretical model is based upon the Landau--de Gennes theory for the nematic phase.
This model is a Landau expansion for the nematic phase \cite{deGennes1995} with respect to a tensor order parameter which enables simulation of phase transition and orientational defects (disclinations).
Surface anchoring is modelled using a Landau-type expansion for the surface free energy \cite[Chap.~4]{Barbero2005}, as well.
The governing equations are solved using the finite element method with second order implicit adaptive time-stepping \cite{FEniCSBook}.
Finally, visualization of the resulting tensor fields uses hyperstreamlines \cite{Delmarcelle1993} with a recently introduced topological seeding method \cite{Fu2015}.

\subsection{Landau--de Gennes Model}

The Landau--de Gennes model for the nematic free energy \cite{deGennes1995} is expanded in terms of a second-order symmetric traceless tensor order parameter which is related to the anisotropic part of the magnetic susceptibility tensor \cite{deGennes1995}.
This tensor is known as the Q-tensor and is defined as:
\begin{equation}\label{eqn:q_tensor}
\bm{Q} = \lambda_{1} \bm{nn} + \lambda_{2} \bm{mm} + \lambda_{3} \bm{ll} =  S \left( \bm{nn} - \frac{1}{3}\bm{\delta} \right) + P\left({ \bm{mm}-\bm{ll} }\right)
\end{equation}
where $\bm{n}$/$\bm{m}$/$\bm{l}$ are eigenvectors of $\bm{Q}$ which characterize the axes of molecular orientation, $\lambda_{i}$ are the eigenvalues of $\bm{Q}$, $S=\frac{3}{2}\lambda_{1}$ is the uniaxial scalar nematic order parameter, and $P = \frac{1}{2}(\lambda_{2}-\lambda_{3})$ is the biaxial scalar nematic order parameter.
$S$ and $P$ quantify the extent to which the molecules conform to the local orientation where $S = P = 0$ in the isotropic phase and $S \ne 0$, $P = 0$ in the uniaxial nematic phase.
However, certain boundary conditions/geometries, external fields, and the presence of disclination defects can result in simultaneous order in multiple directions, resulting in biaxial order where both $S$ and $P$ are non-zero.
 
The Landau--de Gennes model is able to capture both phase transition and elasticity through including terms in the expansion involving both $\bm{Q}$ and $\nabla\bm{Q}$:
\begin{equation}
f_{b}(\bm{Q}, \bm{\nabla Q}, \bm{E}, T) = f_{iso} + f_{nb}(\bm{Q}, T) + f_{ne}(\bm{Q}, \bm{\nabla Q}) + f_{e}(\bm{Q}, \bm{E})
\end{equation}
where $f_{b}$ is the Helmholtz free energy density of the domain, $f_{iso}$ is the free energy contribution of the isotropic phase, $f_{nb}$ is the free energy contribution of bulk nematic ordering (thermodynamic), $f_{ne}$ is the free energy contribution of the nematic distortions (elastic), $f_{e}$ is the free energy contribution from the electric field, and $\bm{E}$ is the electric field vector.
The bulk nematic contribution is (Einstein notation):
\begin{equation}
f_{nb} = \frac{1}{2} a_{0} (T - T_{ni})(Q_{ij} Q_{ji}) - \frac{1}{3} b (Q_{ij} Q_{jk} ) Q_{ki} + \frac{1}{4} c (Q_{ij} Q_{ji})^{2}
\end{equation}
where $a_{0}$/$b$/$c$ are material parameters and $T_{ni}$ is the theoretical second order isotropic/nematic transition temperature.
The nematic elastic contribution is \cite{Barbero2000}:
\begin{equation}
f_{el} = \frac{1}{2} L_{1} \left( \partial_{i} Q_{jk} \partial_{i} Q_{kj}  \right) + \frac{1}{2} L_{2} \left( \partial _{i} Q_{ij} \partial _{k} Q_{kj}\right) + \frac{1}{2} L_{3} \left(\partial_{ k} Q_{ij} \partial_{j} Q_{ik} \right)
\end{equation}
where $L_{i}$ are elastic material parameters which can be related to the fundamental modes of nematic deformation: splay, twist, and bend.
In this study, the one-constant approximation is employed, making the $L_2$ and $L_3$ terms vanish.
The electric field contribution is \cite{Barbero2000}:
\begin{equation}
f_{e} = -\frac{\epsilon_{\circ}}{8\pi}\left[ \left( \frac{\epsilon_{\parallel}+2\epsilon_{\perp} }{3} \delta_{ij} +( \epsilon_{\parallel} -\epsilon_{\perp} ) Q_{ij} \right) \right] E_{j} E_{i} 
\end{equation}
where $\epsilon_{\parallel}$ and $\epsilon_{\perp}$ are the dielectric constants parallel and perpendicular to the director $\bm{n}$ respectively.
The above free energy formulation may be used to approximate the total free energy functional of a nematic domain in the presence of an electric field:
\begin{equation}\label{eqn:total_free_energy}
    F[\bm{Q}] = \int_{V} f_{b} dV
\end{equation}

In order to simulate dynamics, with the previously mentioned assumptions of neglecting flow and thermal fluctuations, the dynamic equation used corresponds to the time-dependent Ginzburg-Landau model or so-called Model A dynamics \cite{Hohenberg1977}:
\begin{equation}
    \frac{\partial Q_{ij}}{\partial t} = - \Gamma \left[\frac{\delta F}{\delta Q_{ij}}\right]^{ST}
\end{equation}
where $\Gamma = \mu_{r}^{-1}$, $\mu_{r}$ is the rotational viscosity of the nematic phase, and $[]^{ST}$ is the symmetric-traceless component of the expression.
At equilibrium this expression is equal to the Euler-Lagrange equation for the total free energy functional (eqn. \ref{eqn:total_free_energy}).
Expanding the functional derivative in the dynamic governing equation results in \cite{Barbero2000}:
\begin{equation}
    \mu_{r} \frac{\partial Q_{ij}}{\partial t} = \left[\frac{\partial f_{b}}{\partial Q_{ij}} - \partial_{k} \left( \frac{\partial f_{b}}{\partial \left( \partial_{k} Q_{ij}\right) } \right) \right]^{ST}
\end{equation}

\subsection{Surface Anchoring Energy}

The two major types of physical anchoring conditions for nematic interfaces are homeotropic and planar, corresponding to anchoring parallel and orthogonal to the surface normal, respectively.
In this work, homeotropic anchoring is utilized in that it uniquely constrains the preferred anchoring direction, whereas the use of planar anchoring would introduce additional degrees of freedom in the boundary conditions.
As with the bulk nematic domain, a surface free energy density can be formulated in terms of the Q-tensor field and the surface unit normal $\bm{k}$ \cite{Barbero2005}:
\begin{equation}
f_{s} = f_{s,iso} - \alpha k_{i} Q_{ij} k_{j}
\end{equation}
where $f_{s,iso}$ is the isotropic contribution to the surface free energy, $\alpha$ is the anchoring strength, and only terms up to first-order are retained.
The total free energy (eqn. \ref{eqn:total_free_energy}) must now include a contribution from the bounding surface:
\begin{equation}
    F[\bm{Q}] = \int_{V} f_{b} dV + \int_{S} f_{s} d S
\end{equation}
which imposes a boundary condition on the governing dynamic equation:
\begin{equation}
\frac{\partial \gamma}{\partial Q_{ij}}+ k_{k} \frac{\partial f_{b}}{\partial \left( \partial_{k} Q_{ij}\right) } = 0
\end{equation}

\subsection{Numerical Methods and Simulation Conditions}

Simulations were performed in two-dimensional elliptic cylinder geometries using the method of lines, where spatial discretization is achieved using the finite element method \cite{FEniCSBook} and time-stepping through a time-adaptive second-order implicit method.
Mesh-independence simulations were first performed, finding that a uniformly distributed node density of \SI{2.06e5}{nodes/\micro\metre^2} was sufficient using an error tolerance of $10^{-10}$.
Verification of simulation results was performed through convergence tests of equilibrium solutions.
Simulation results were visualized using hyperstreamlines \cite{Delmarcelle1993} in conjunction with a topologically-informed seeding method for alignment tensor fields \cite{Fu2015}.
Material parameters were used that approximate pentyl-cyanobiphenyl (5CB) \cite{Coles1979a,Wincure2007a} shown in Table \ref{tab:model_parameters}.

\begin{table}
    \caption{Material parameters for 5CB.\label{tab:model_parameters}}
    \begin{ruledtabular}
    \begin{tabular}{ccc}
    $T_{ni}$ & $307.2$ &  \si{K} \\
    $a_{0}$ & $ 1.4\times 10^{5}$ & \si{J/m^3 K} \\
    $b$ & $1.8\times 10^{6}$ & \si{J/m^3} \\
    $c$ & $3.6\times 10^{6}$ & \si{J/m^3} \\
    $L_{1}$ & $3.0\times 10^{-12}$ & \si{J/m} \\
    $\epsilon_{\parallel}$ & 17 &  (relative) \\
    $\epsilon_{\perp}$ & 7 &  (relative) \\
    \end{tabular}
    \end{ruledtabular}
\end{table}

Two types of simulations were performed: formation dynamics and switching dynamics.
Formation dynamics simulations correspond to cooling of the PDLC domains from isotropic to nematic in the absence of an electric field; starting above the bulk nematic transition temperature (\SI{308.4}{\kelvin}) and cooling below to \SI{307}{\kelvin}. Heterogeneous nucleation of the nematic phase was assumed to be the dominant nucleation mechanism based on recent experimental observations \cite{Aya2011}.
The initial conditions for these simulations assume a boundary layer that is uniaxial and well-aligned with the preferred orientational axis of the surface $\bm{k}(\theta )$, where $\theta$ is the polar angle of each surface point (see Appendix \ref{sec:app1}).
Switching dynamics simulations correspond to application of an electric field to a fully-formed nematic domain.  Simulations were performed with varying electric field strengths and with the field oriented along the major axis of the elliptic cylinder domain. The nondimensionalized form of the governing equation was used, and thus time is reported here as a dimensionless quantity,

\begin{equation}
     \tilde{t} =\frac{t}{\tau }\quad ,\quad \tau =\frac{\mu_r}{a_{0}T_{ni}}
\end{equation}

%%%%%%%%%%%%%%%%%%%%%%%%%%%%%%%%%%%%%%%%%%%%%%%%%%%%%%%%%%%%%%%%%%%%%%%%%%%%%%%%%%%%%%%
\section{Results and Discussion}

Characteristic length scales were used to determine physically relevant domain sizes, surface anchoring energy, and electric field strengths for simulations.
The most fundamental length scale is that of nematic ordering itself, which is determined through the competition between bulk thermodynamic and elastic contributions to the free energy \cite{deGennes1995}:
\begin{equation}\label{eqn:coherence_length}
    \lambda_{n} = \sqrt{\frac{L_{1}}{a_{0}\left(T - T_{ni} \right)}}
\end{equation}
which is the \emph{nematic coherence length}.
This length scale approximates the thickness of the nematic/isotropic interface.
The nematic coherence length is typically $\lambda_{n} \approx \SI{4}{nm}$ in the majority of LCs used in display technology.  

The nematic domain size was chosen to be on the order of $\SI{1}{\micro\metre}$, much larger than $\lambda_{n}$, with a constant area of $\SI{0.8}{\micro\meter^{2}}$ for all simulation geometries.
For elliptic cylinder domains this constrains the cross-sectional area such that $\pi ab = \SI{0.8}{\micro\meter^{2}}$ where $a$ is the semi-major axis and $b$ is the semi-minor axis of the ellipse.
The $\si{\micro\meter}$-scale for LC domains is relevant to PDLCs used for privacy glass and other scattering applications \cite{Drzaic1995}.
Spheroidal geometries have been observed by Drzaic and others \cite{Drzaic1988} that are relatively complex, but range in aspect ratio from approximately $R = (1, 2]$; thus this range of aspect ratios was used in simulations.

Another characteristic length quantifies the competition between the electric field and the nematic elastic forces:
\begin{equation}
    \lambda_{e} = \sqrt{\frac{L_{1}}{(\epsilon_{\parallel} - \epsilon_{\perp}) E^{2}}}
\end{equation}
which is the \emph{dielectric coherence length} \cite{Barbero2005}.
The range of electric field strengths that was studied was $[\SI{0}{\volt\per\micro\meter},\SI{5}{\volt\per\micro\meter}]$ which are typical field strengths for PDLC devices \cite{Bronnikov2013}.
These field strengths result in $\lambda_{e} \rightarrow \SI{10}{\nano\meter}$ which corresponds to a moderately strong electric field, but not to the extent that it could induce melting of the nematic phase.

A similar characteristic length can be determined for the competition between surface anchoring and elastic forces:
\begin{equation}
    \lambda_{s} = \frac{L_{1}}{ \alpha}
\end{equation} 
which is the \emph{surface extrapolation length} \cite{Barbero2005}.
As $\lambda_{s} \rightarrow 0$ the surface anchoring effects dominate and the nematic alignment at the boundary governs the bulk texture; this is so-called ``strong'' anchoring.
As $\lambda_{s} \rightarrow \infty$ the bulk nematic elasticity effects dominate and the nematic alignment at the surface is governed by the bulk texture.
A value for the surface anchoring strength of \SI{5e-5}{\joule\per\meter^{2}} was used with $\lambda_{s} \approx \SI{100}{\nano\meter}$ which corresponds to ``weak'' surface anchoring.

\subsection{Nematic Domain Formation and Equilibrium Texture}\label{sec:formation}

\begin{figure*}
    \begin{subfigure}[b]{\linewidth}
        \centering
        \includegraphics[width=0.2\linewidth]{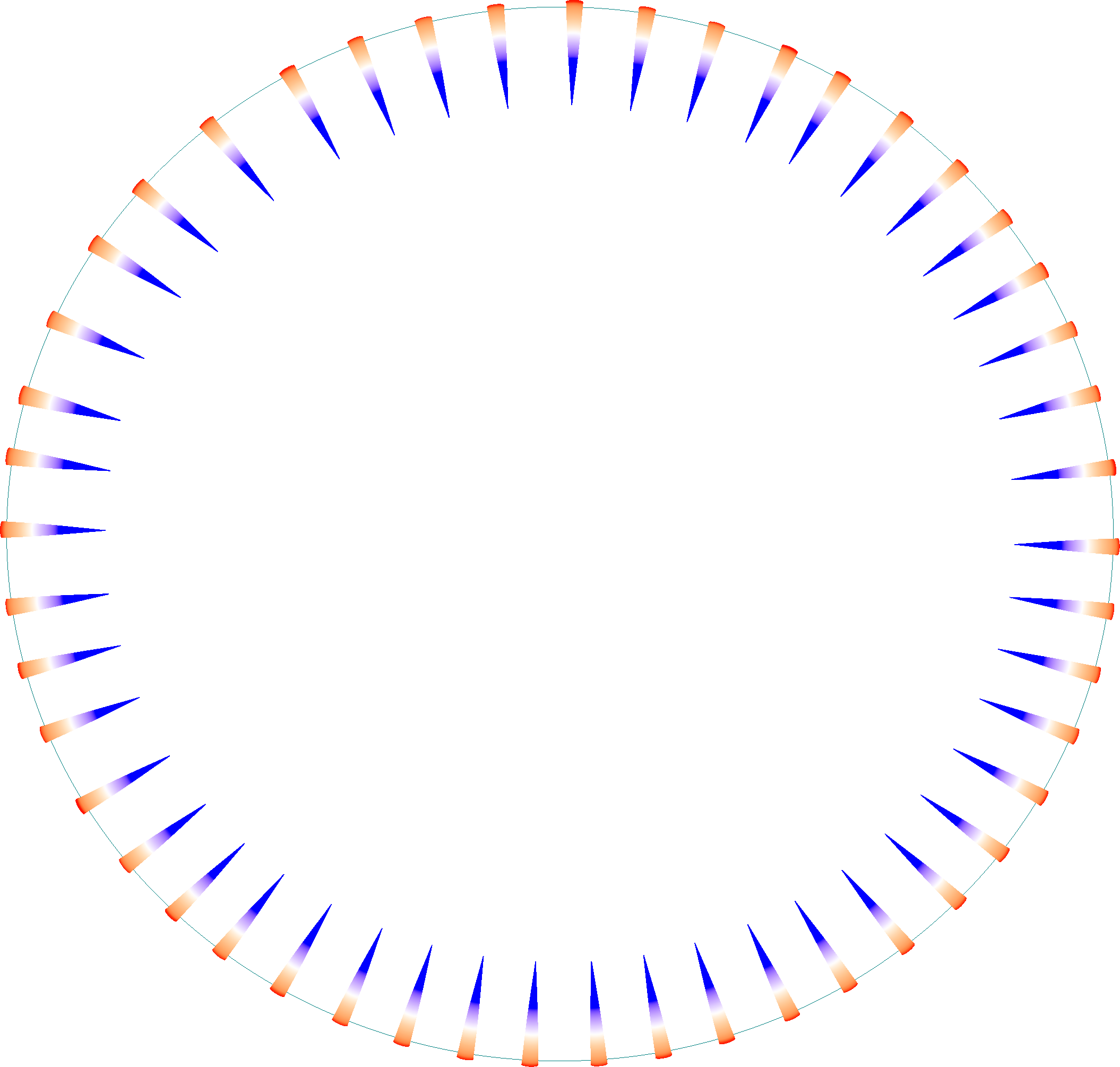} %t = 50
        \quad
        \includegraphics[width=0.2\linewidth]{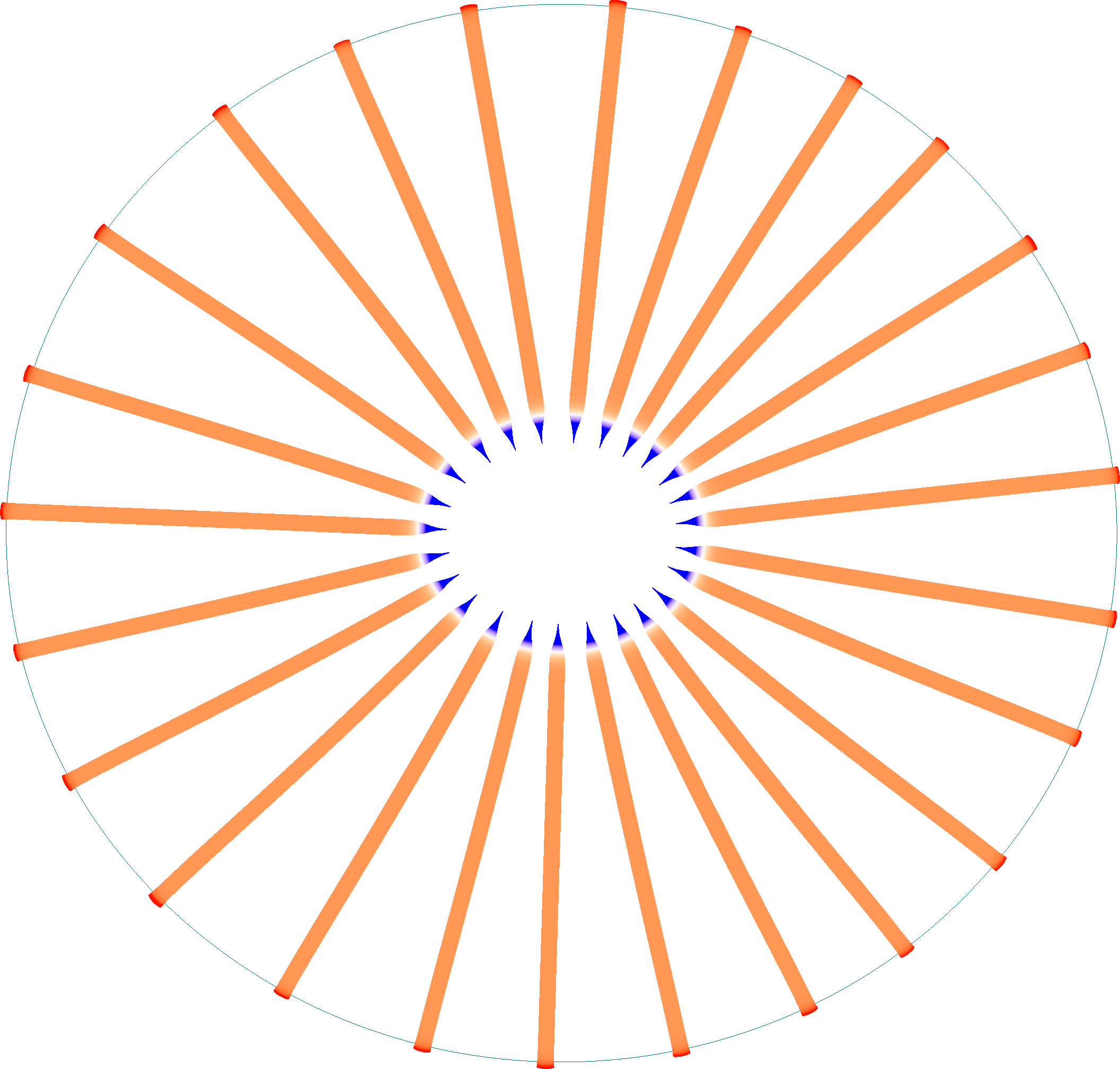} %t = 18233
        \quad
        \includegraphics[width=0.2\linewidth]{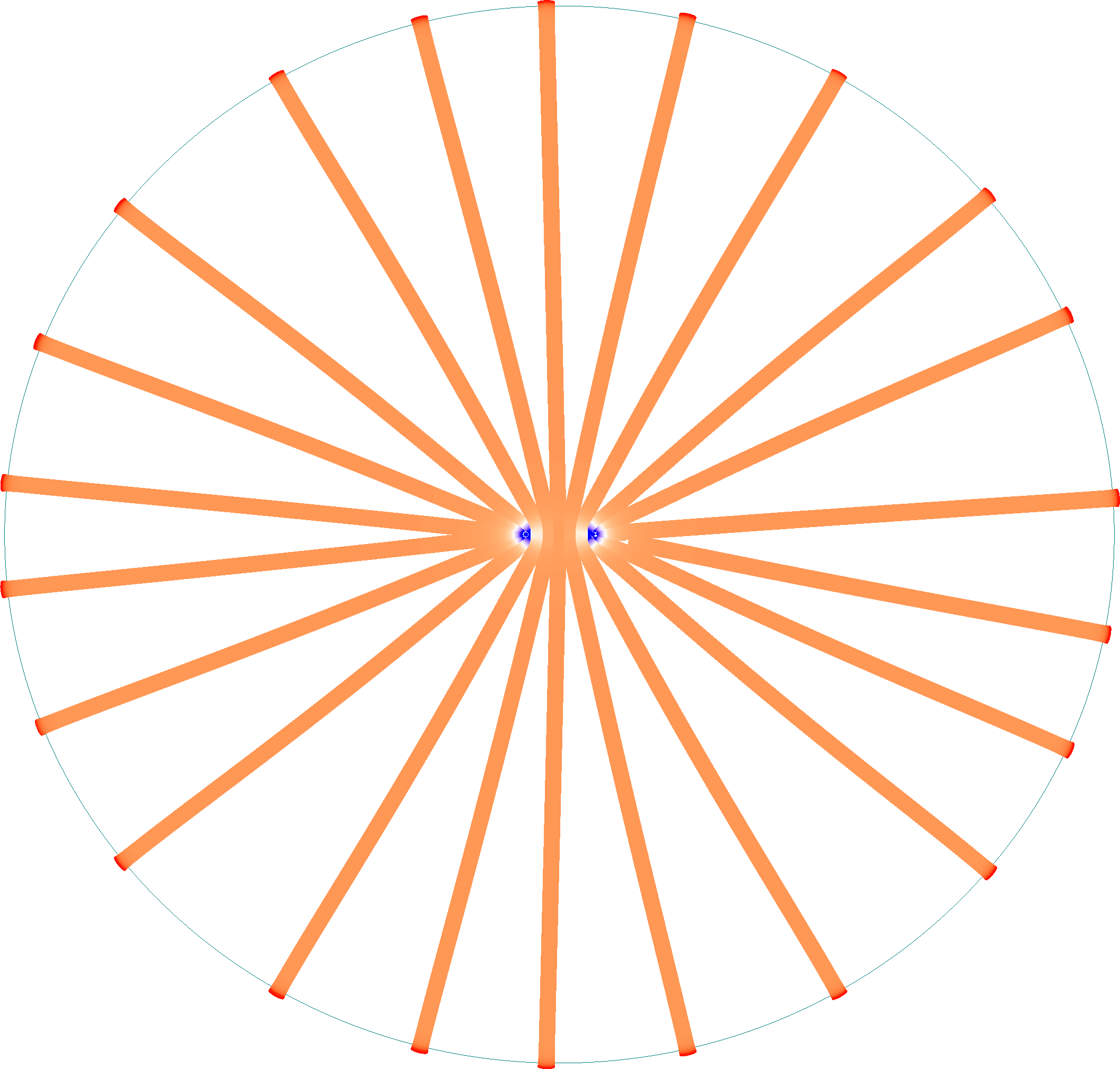} %t = 24314
        \\
        \includegraphics[width=0.2\linewidth]{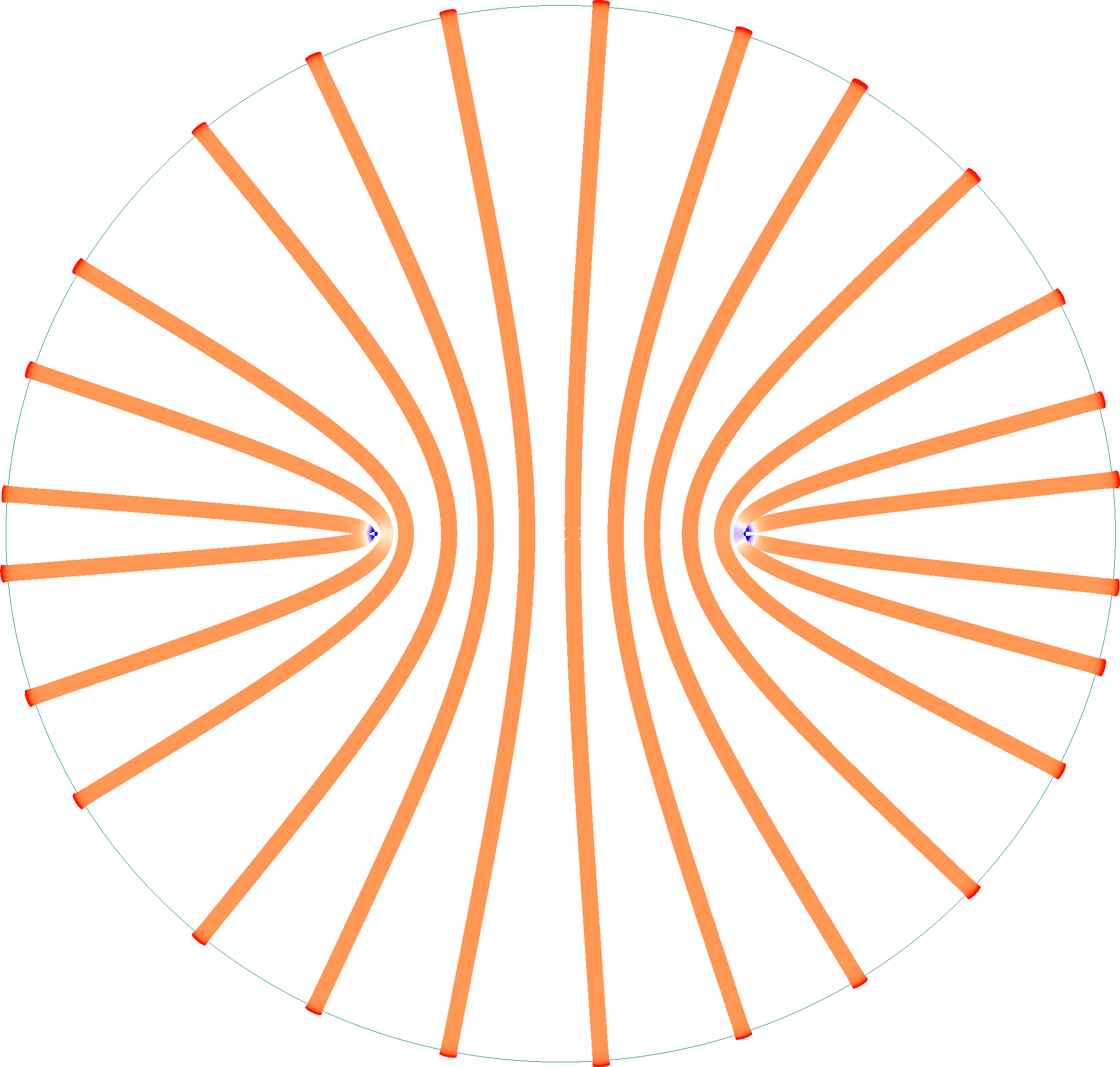} %t = 462713
        \quad
        \includegraphics[width=0.2\linewidth]{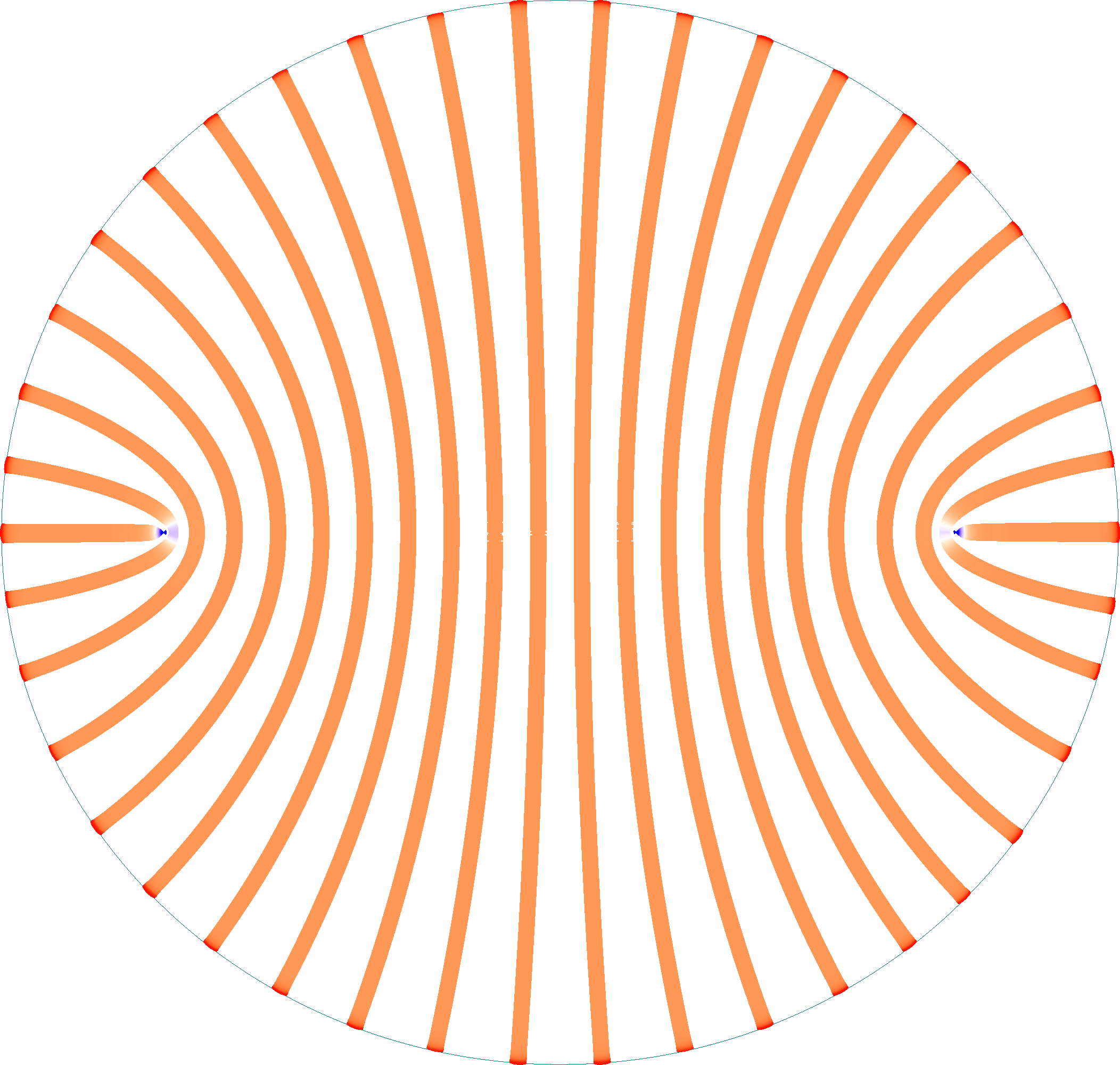} %t >= 6401934
        \caption{}\label{fig:formation_10}
    \end{subfigure}\\
    \begin{subfigure}[b]{\linewidth}
        \centering
        \includegraphics[width=0.25\linewidth]{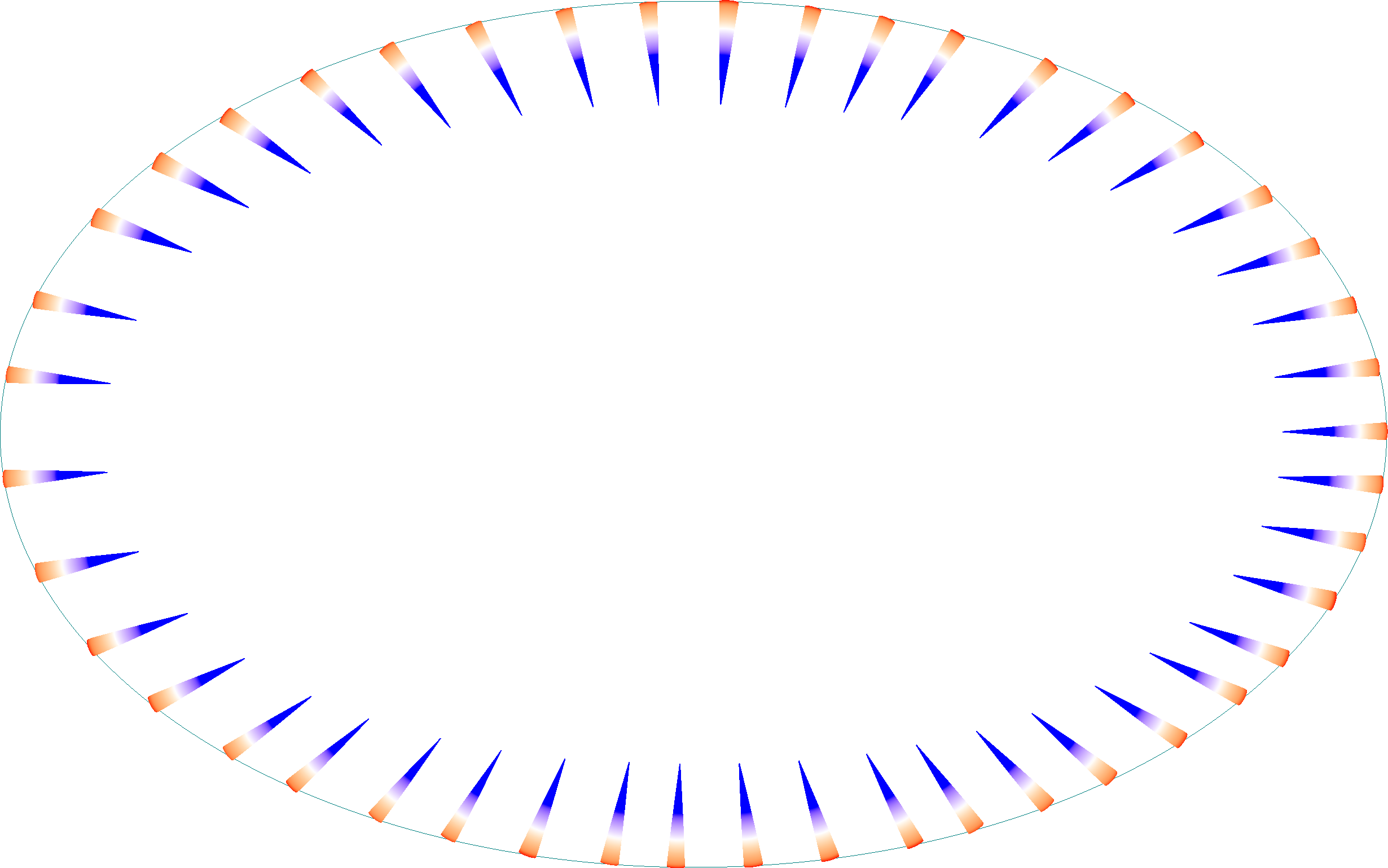} %t = 50
        \quad
        \includegraphics[width=0.25\linewidth]{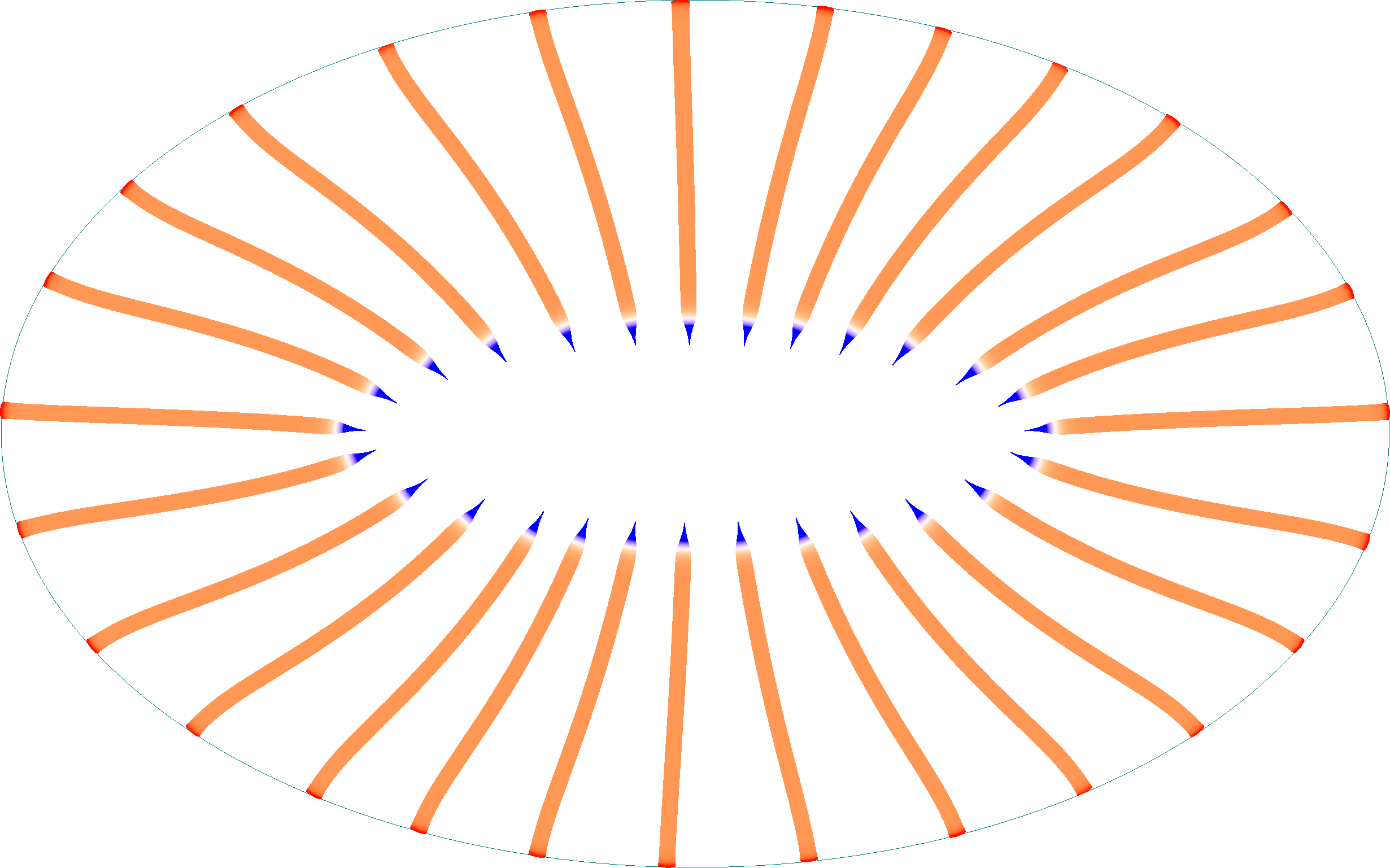} %t = 13473
        \quad
        \includegraphics[width=0.25\linewidth]{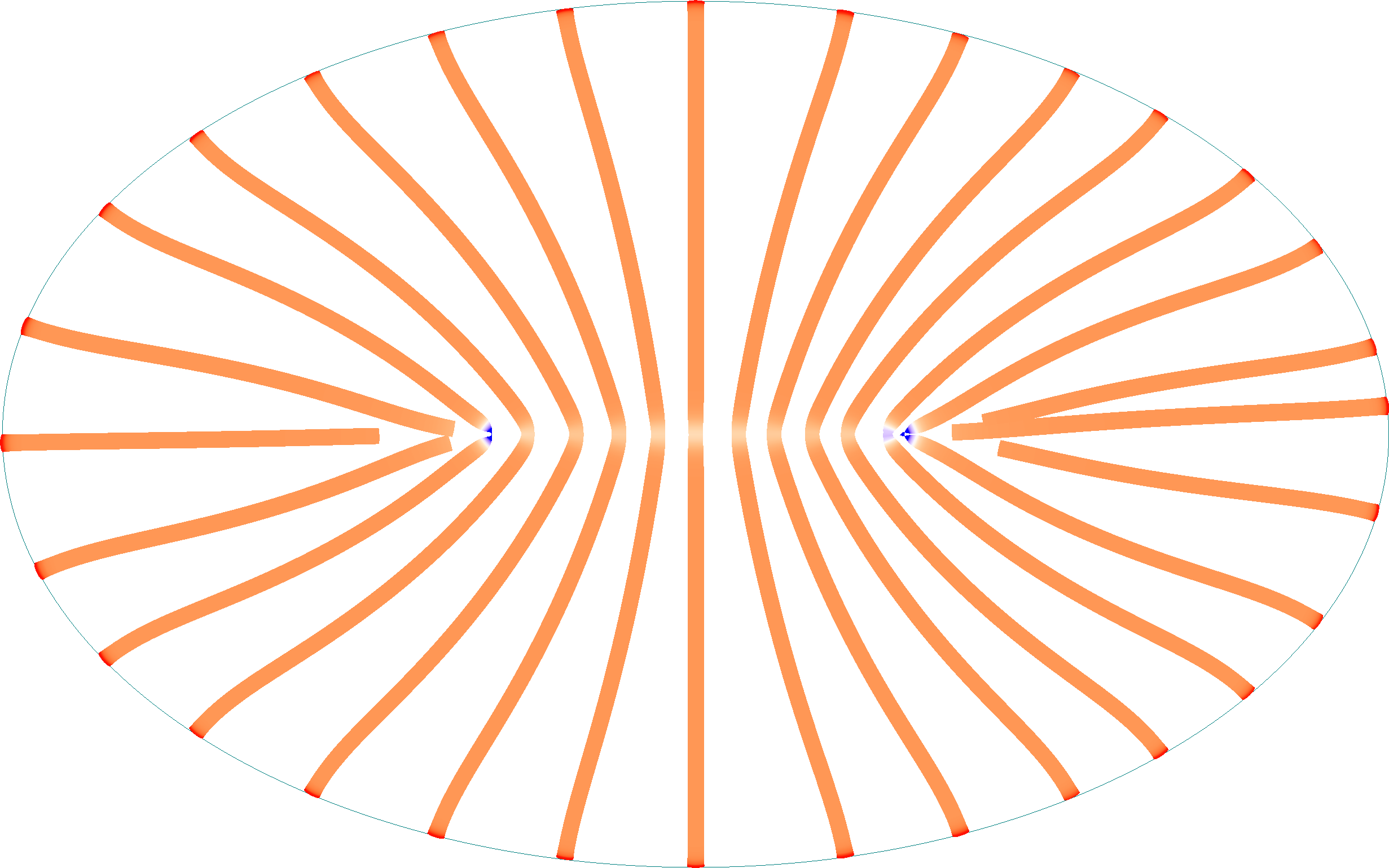} %t = 19283
        \\
        \includegraphics[width=0.25\linewidth]{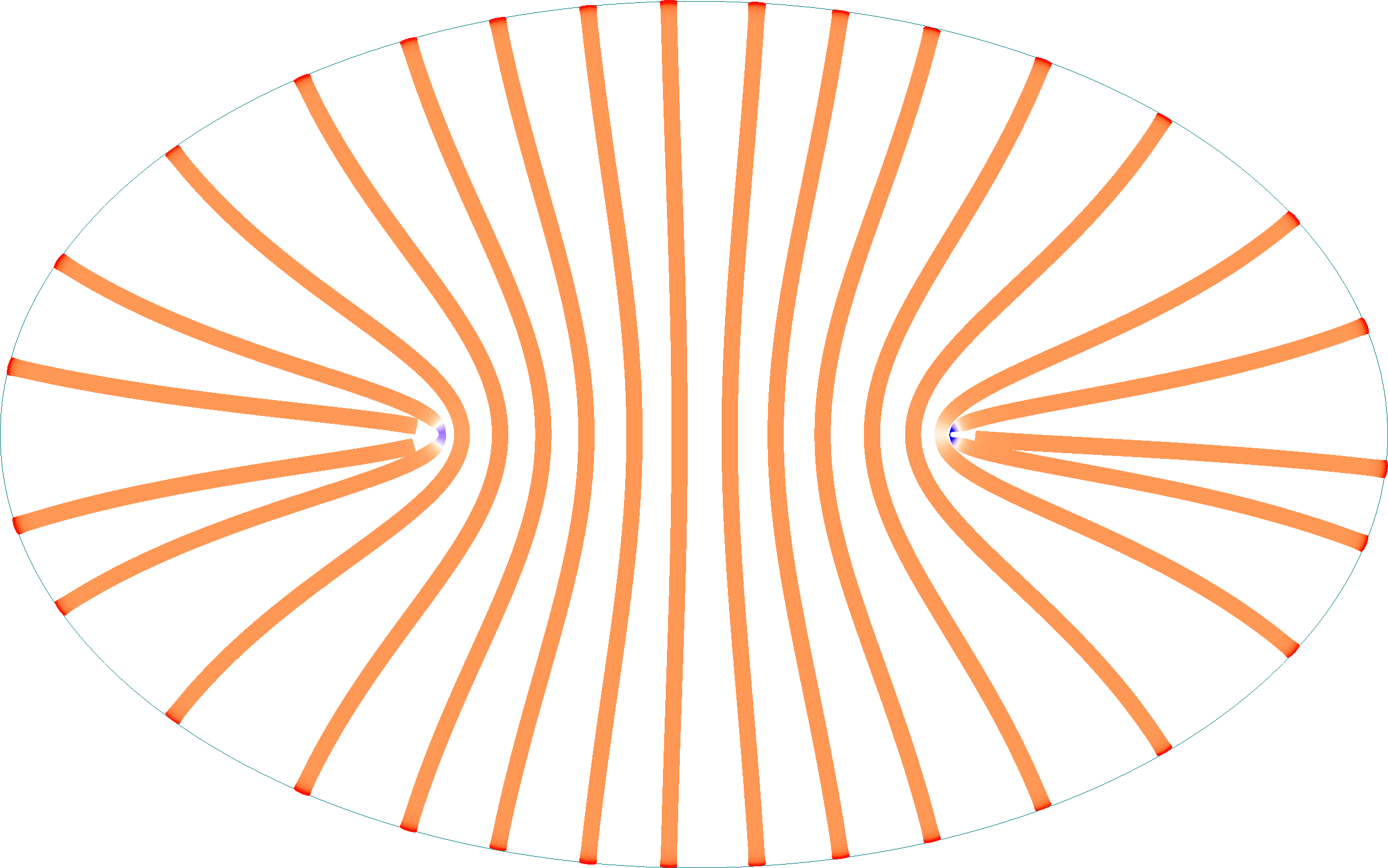} %t = 401801
        \quad
        \includegraphics[width=0.25\linewidth]{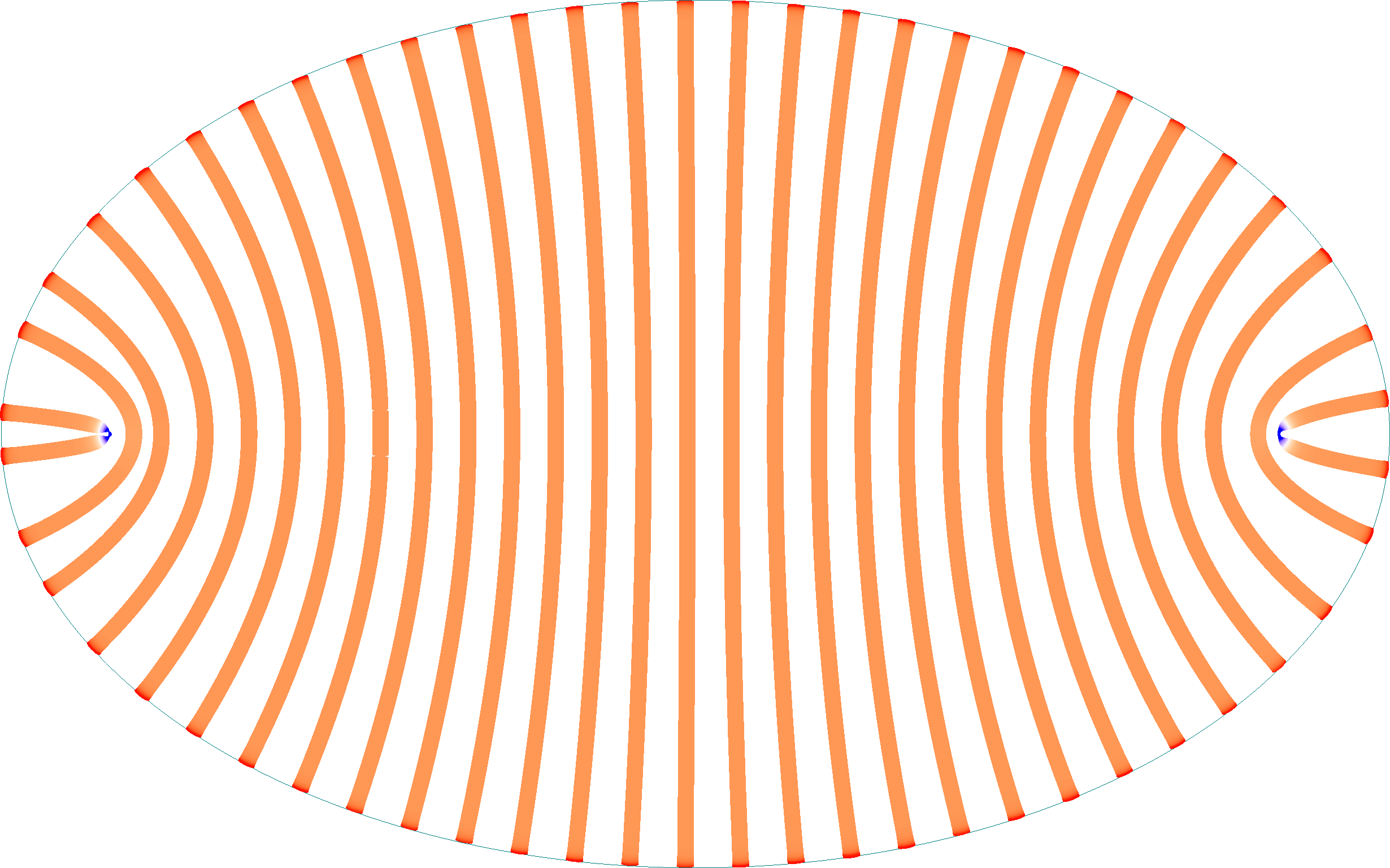} %t > 4564292
        \caption{}\label{fig:formation_16}
    \end{subfigure}\\
    \begin{subfigure}[b]{\linewidth}
        \centering
        \includegraphics[width=0.3\linewidth]{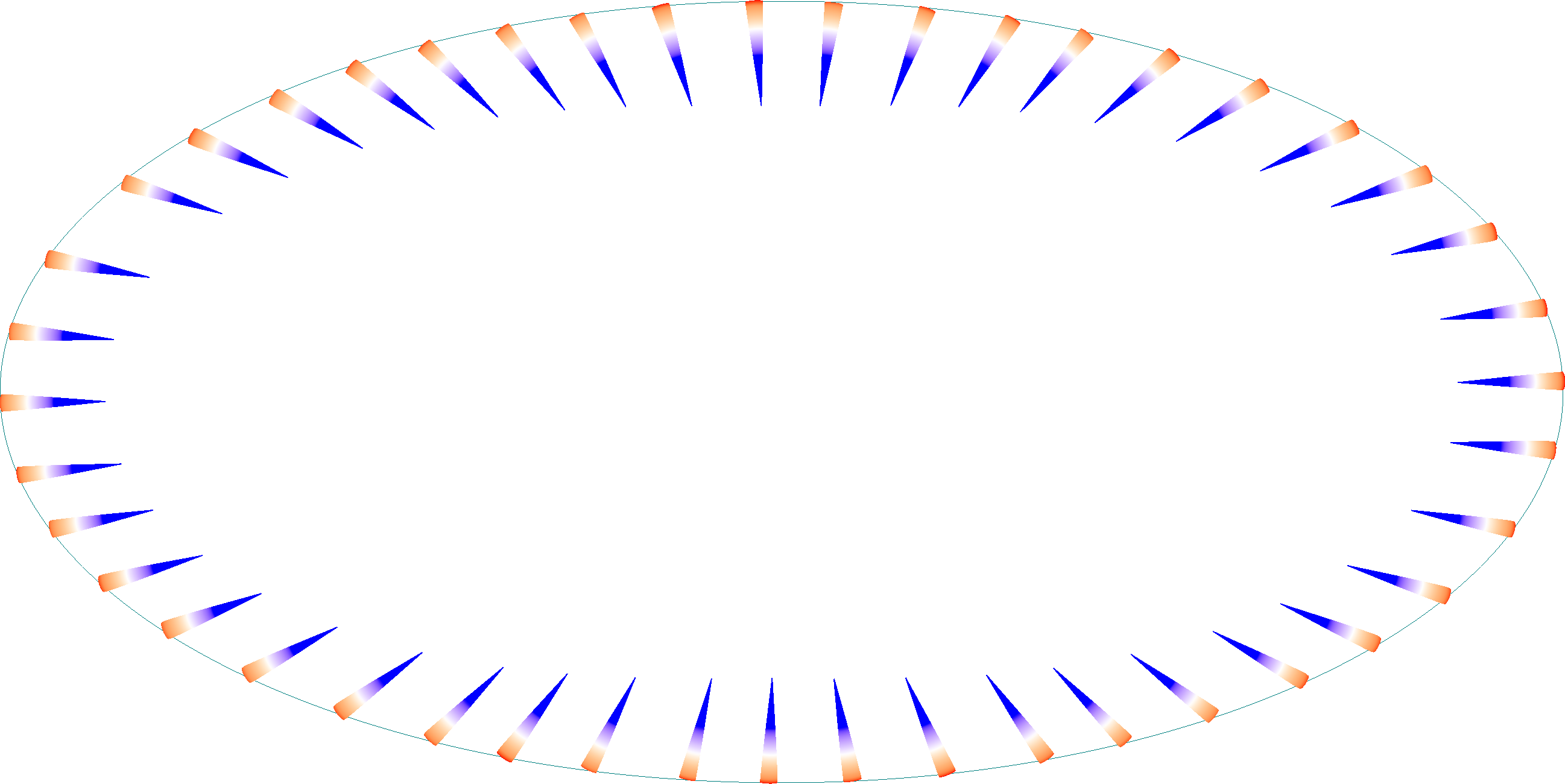} %t = 50
        \quad
        \includegraphics[width=0.3\linewidth]{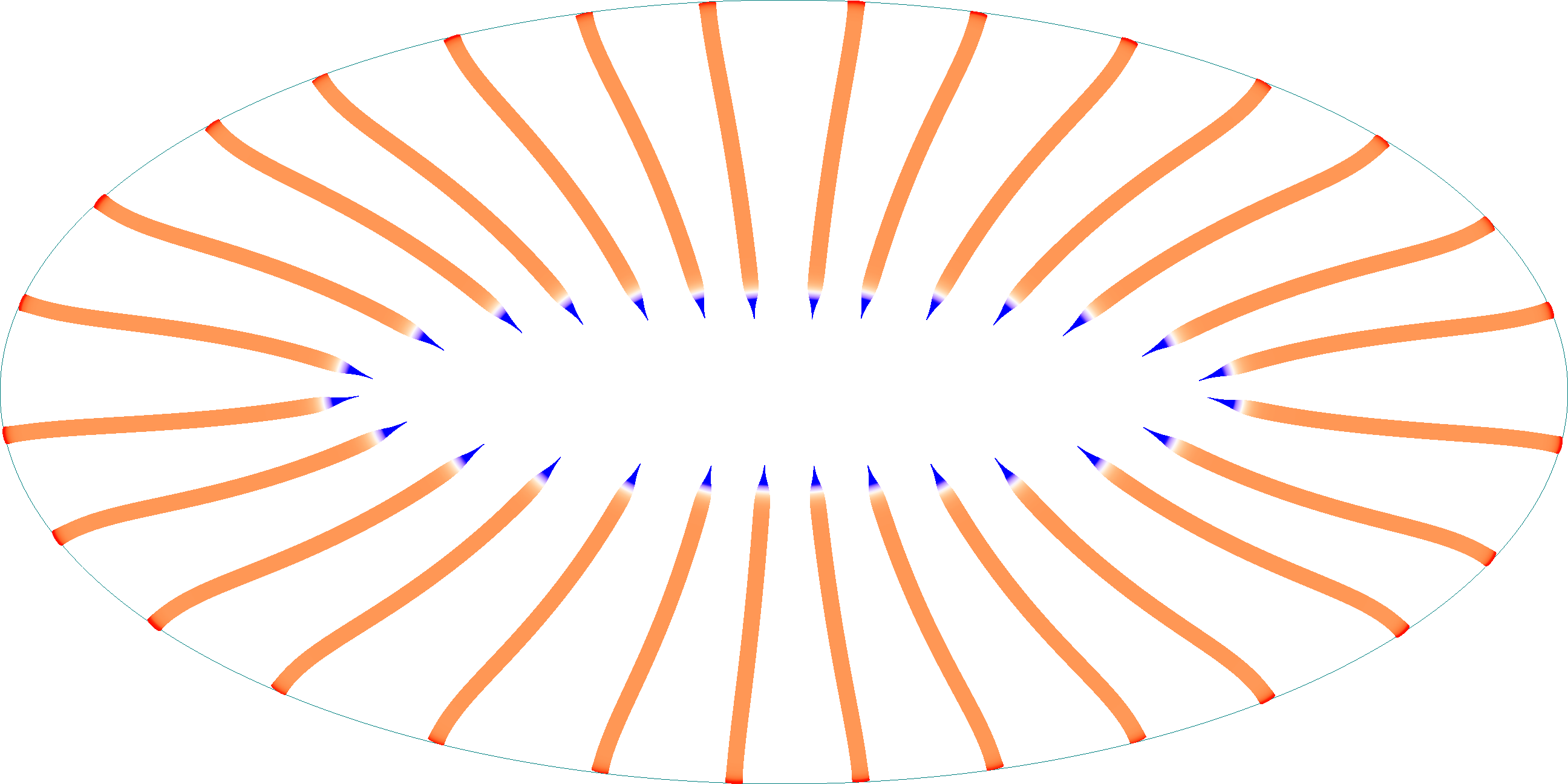} %t = 11924.3
        \quad
        \includegraphics[width=0.3\linewidth]{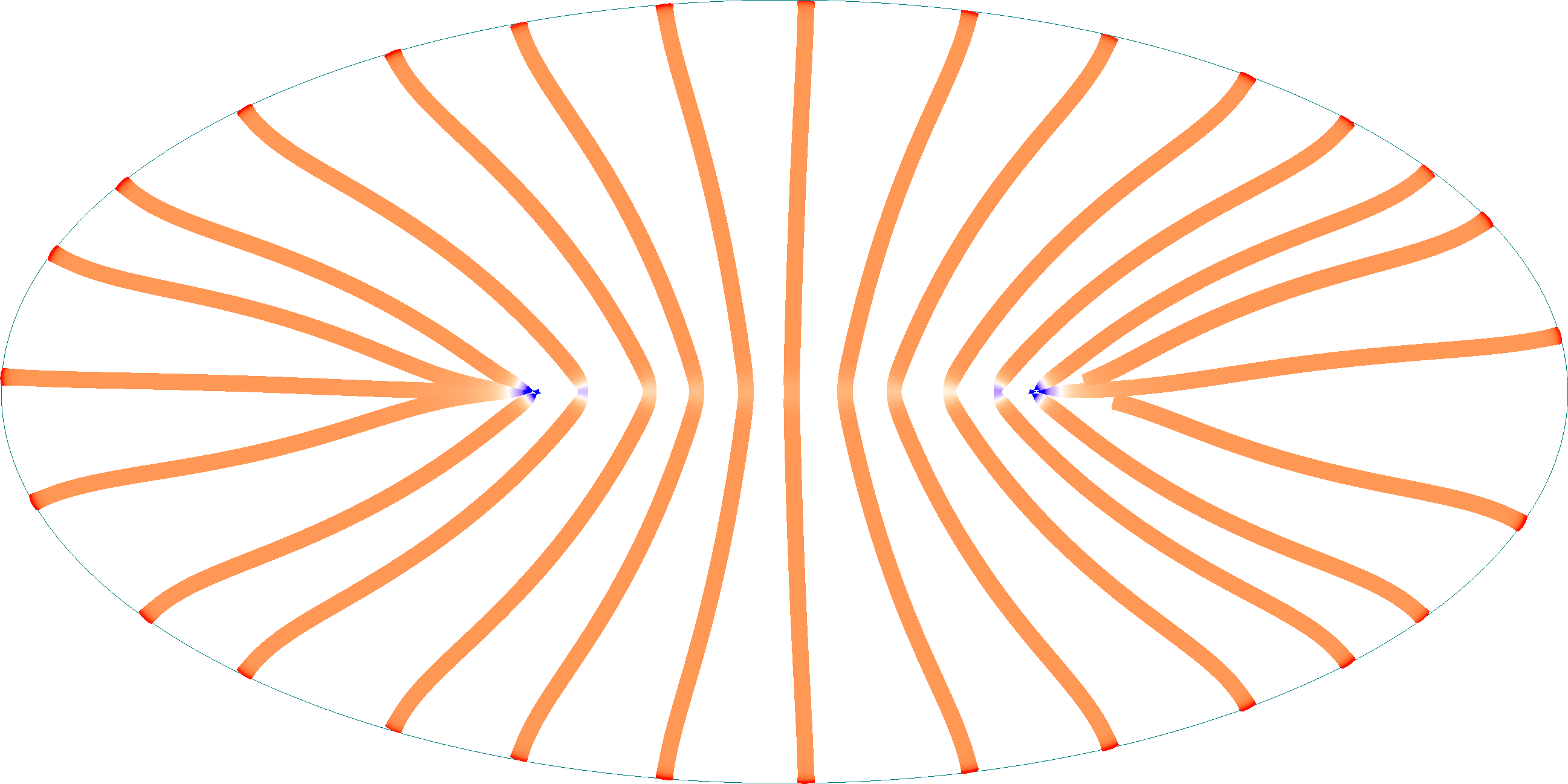} %t = 17058.1
        \\
        \includegraphics[width=0.3\linewidth]{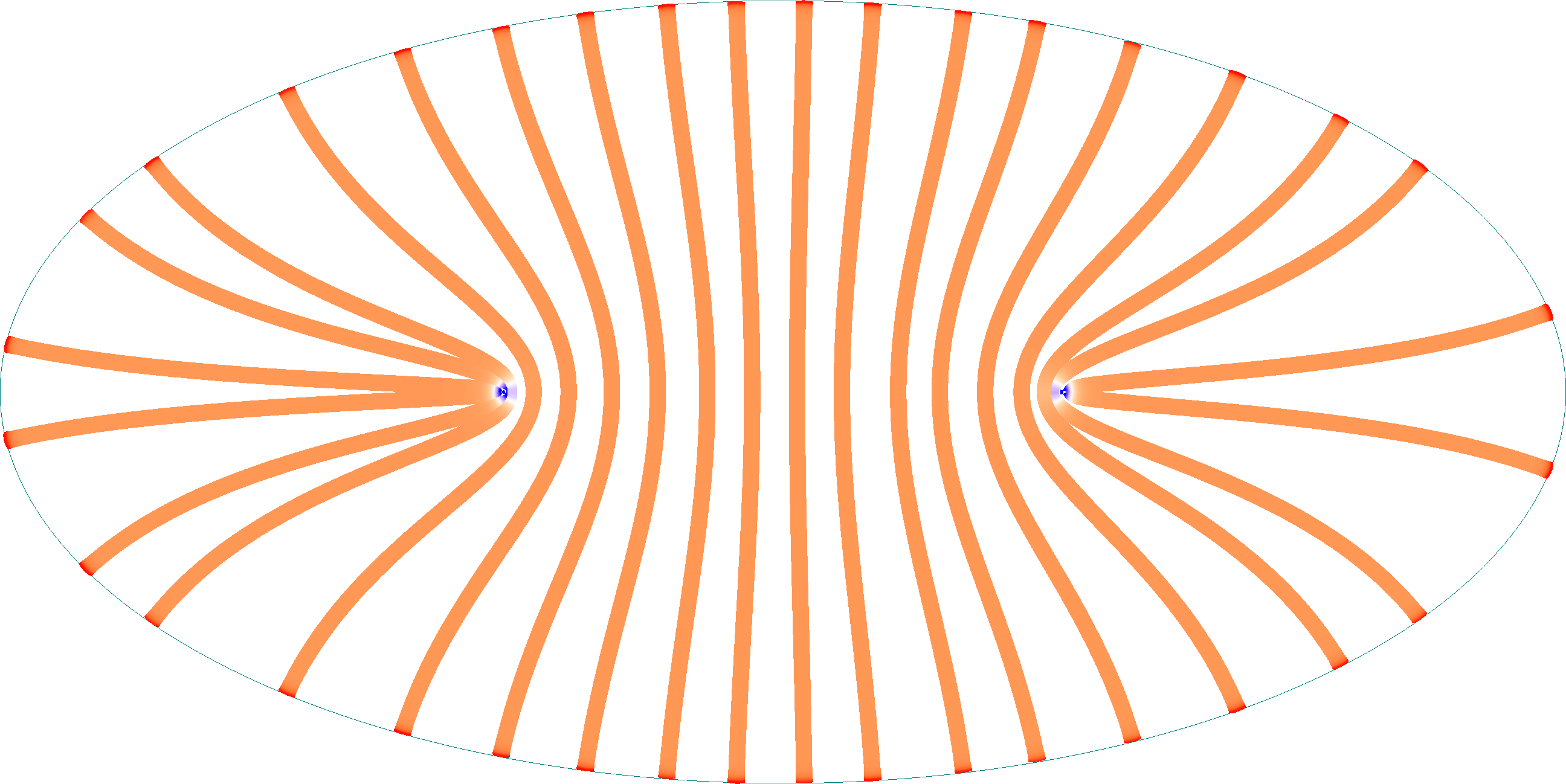} %t = 309083
        \quad
        \includegraphics[width=0.3\linewidth]{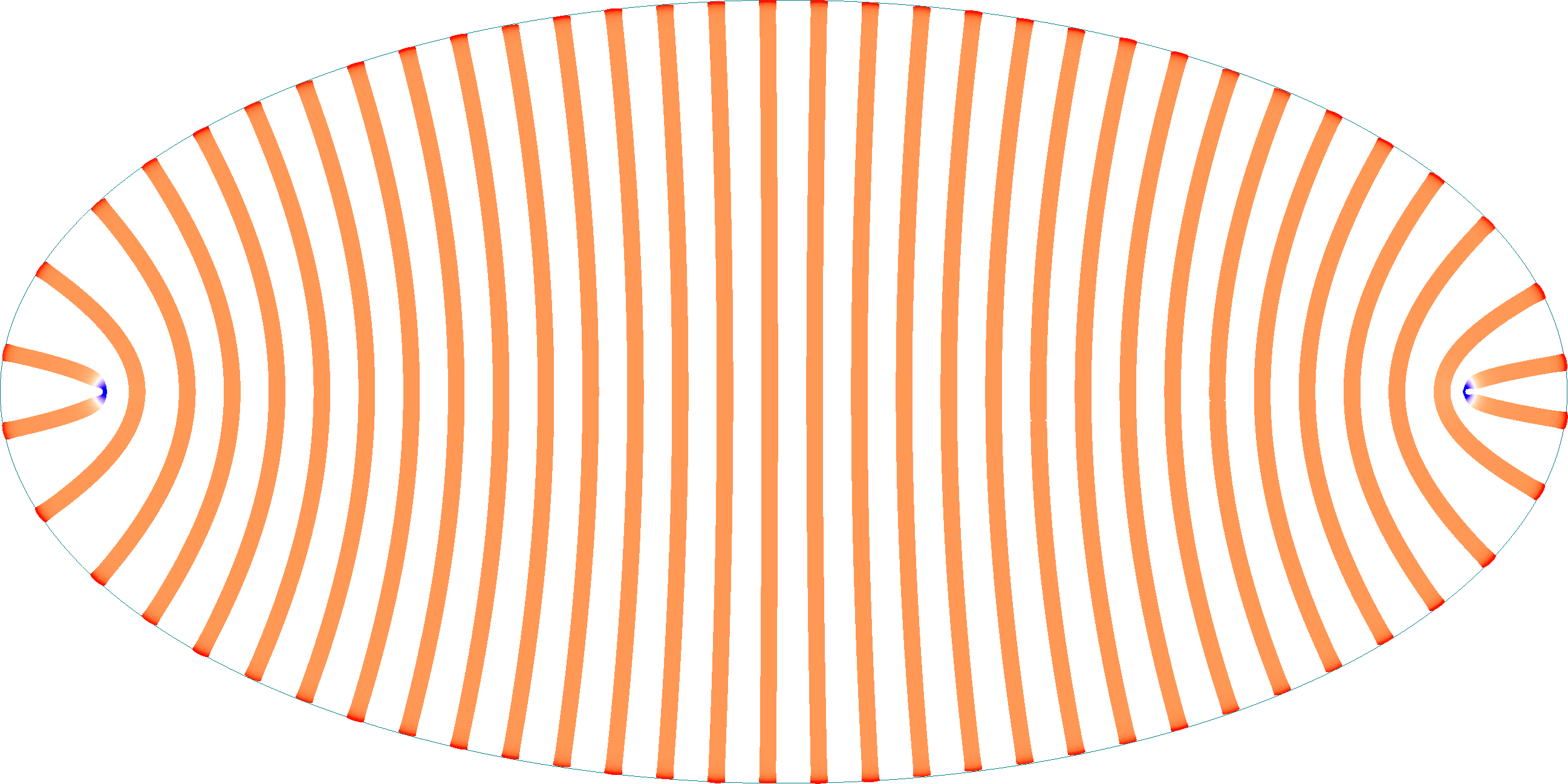} %t >= 3348801
        \caption{}\label{fig:formation_20}
    \end{subfigure}
    \includegraphics[scale=0.07]{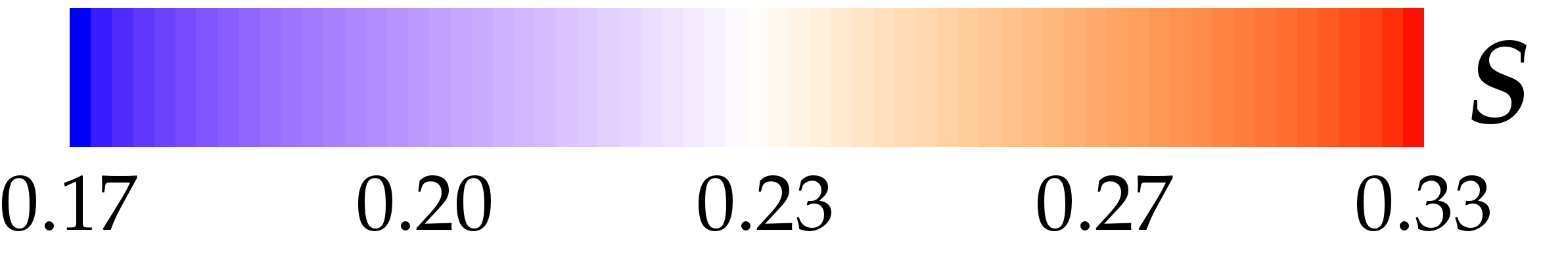}
    \caption{(Color online) Hyperstreamline visualizations of the Q-tensor fields during the formation process for nematic domains: (a) $R=1.05$ with $t = \SI{50}{}$, $\SI{1.82e4}{}$, $\SI{2.43e4}{}$, $\SI{4.63e5}{}$, $\SI{6.40e6}{}$; (b) $R=1.6$ with $t = \SI{50}{}$, $\SI{1.35e4}{}$, $\SI{1.93e4}{}$, $\SI{4.02e5}{}$, $\SI{4.56e6}{}$; and (c) $R=2$ with $t = \SI{50}{}$, $\SI{1.19e4}{}$, $\SI{1.71e4}{}$, $\SI{3.09e5}{}$, $\SI{3.35e6}{}$. }
    \label{fig:formation}
\end{figure*}

Fig.~\ref{fig:formation} shows visualizations of the Q-tensor field for nematic elliptic cylinder domains with aspect ratios $R=\{1.05,1.6, 2\}$.
The dynamics observed in these simulations are representative of those for all aspect ratios studied, $R=(1,2]$.
A sequence of three distinct growth regimes was observed: free growth, interface impingement/defect formation, and bulk relaxation.
The \emph{free growth regime} involves the stable nematic phase growing into the unstable isotropic phase such that the texture is approximately commensurate with the anchoring conditions.
The \emph{interface impingement/defect formation regime} follows, where the nematic/isotropic interface impinges on itself resulting in the simultaneous formation of a pair of orientational defects along the major axis.
For all simulations, the type of orientational defects, or disclinations, observed were wedge-type with strength $+\frac{1}{2}$ \cite[Chap.~2]{Kleman1982}.
These disclinations are formed in order to resolve the topological constraints imposed by the confinement geometry and anchoring conditions.
Finally, the \emph{bulk relaxation regime} follows impingement where the fully formed nematic texture relaxes to its equilibrium state through simultaneous disclination motion towards the ellipse focal regions and bulk reorientation.

The growth and formation mechanism for the $R \approx 1$ case has been studied in past work for two-dimensional domains \cite{Sonnet1995,Kralj1995,Yan2002,Sharma2003}.
Rey and Sharma developed a texture phase diagram using the Landau--de Gennes model for the circular geometry \cite{Sharma2003}.
They predict the splitting of a $+1$ disclination into a pair of $+\frac{1}{2}$ disclinations located at the center of the domain for domain sizes on the order of $\lambda_{n}$.
Given that the simulated domains are micron-scale, the defect splitting observed in the $R \approx 1$ case is in agreement with these past results.
The resulting domain texture is uniformly oriented in a region parallel to an axis and with the pair of disclinations located along the axis orthogonal to it.
Additionally, simulation results are found to also be consistent with past work \cite{Kralj1995} predicting a transition from textures with disclinations to an ``escape'' texture, absent of defects, under certain conditions (anchoring strength, domain size, etc).
The growth and formation mechanism for the elliptic case $R > 1$ is observed to be significantly different and does not exhibit disclination splitting, even though the topologies of the elliptic and circular domains are equivalent.

The underlying formation mechanism for elliptic domains is well-explained by the approximation of Wincure and Rey \cite{Wincure2006} of the velocity of a uniaxial isotropic/nematic interface.
They have shown that its velocity $v$ is proportional to the difference in energy between the nematic and isotropic phases $\Delta F$ and capillary force $\mathcal{C}$ \cite{Wincure2006}:
\begin{equation}
    \beta v = \mathcal{C} - \Delta F
\end{equation}
where $\beta$ is an effective viscosity term.
During the free growth regime, in every case (Fig.~\ref{fig:formation}), as the isotropic/nematic interface approaches the center of the domain, the capillary force grows inversely proportional to the radius of the central isotropic region.
The first deviation of the elliptic formation mechanism is observed due to this competition of forces.
For a circular domain, $v(\theta)$ is essentially constant but for the elliptic domain it varies from a maximum at the interface regions closest to the minor axis to a minimum at those regions closest to the major axis.

As the radius of curvature of the interfaces closest to the major axis approaches a critical value proportional to the nematic coherence length $\lambda_n$, the capillary force approaches the difference in free energy driving force ($\mathcal{C}\rightarrow \Delta F$).
This results in a critical slowing down of the interface and the transition from the free growth regime to the impingement/defect formation regime.
The conditions under which this occurs for the elliptic case cannot result in the formation of a $+1$ disclination which later splits, as is observed for circular domains \cite{Sharma2003}.
Instead, a pair of $+\frac{1}{2}$ disclinations form directly along the major axis of the elliptic domain near the two (ellipse) focal points.
Simultaneously, the isotropic/nematic interfaces in the central region impinge, forming a well-aligned central region along the minor-axis of the elliptic domain.
This defect formation mechanism is the confinement-driven analogue to the defect ``shedding'' mechanism discovered by Wincure and Rey for free growth of nematic droplets in an isotropic matrix phase \cite{Wincure2007a}.
In both cases, defects are formed at the isotropic/nematic interface due to frustration between bulk droplet texture and interfacial anchoring.

\begin{figure}[h!]
        \includegraphics[width=0.45\linewidth]{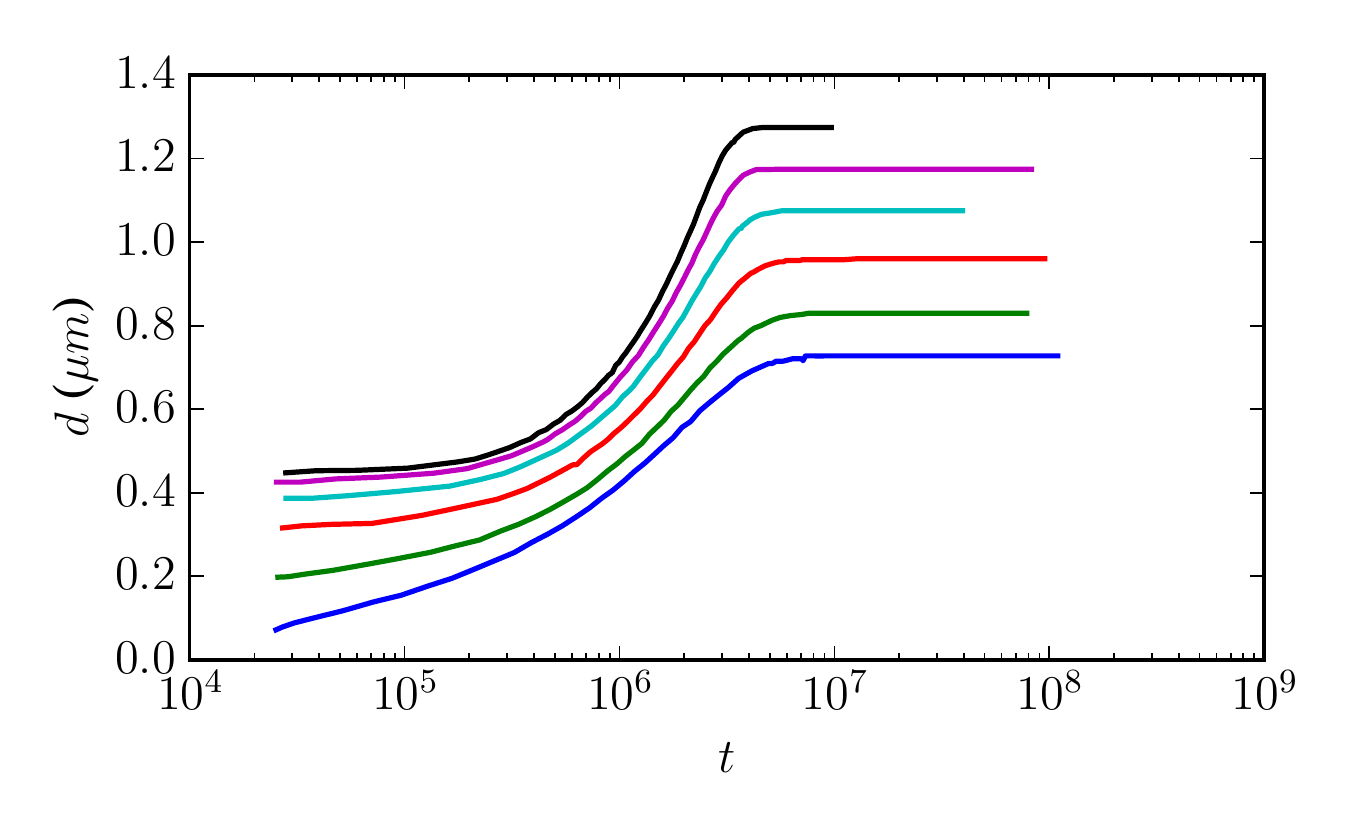}
    \caption{(Color online) Evolution of the distance $d$ between defects for different aspect ratio equilibrium elliptic domains. In order from bottom curve to top, the aspect ratio is $R = \{1.05, 1.2, 1.4, 1.6, 1.8, 2.0\}$. \label{fig:equilibrium_distance}}
\end{figure}

\begin{figure}[h!]
    \begin{subfigure}[b]{0.45\linewidth}
        \includegraphics[width=\linewidth]{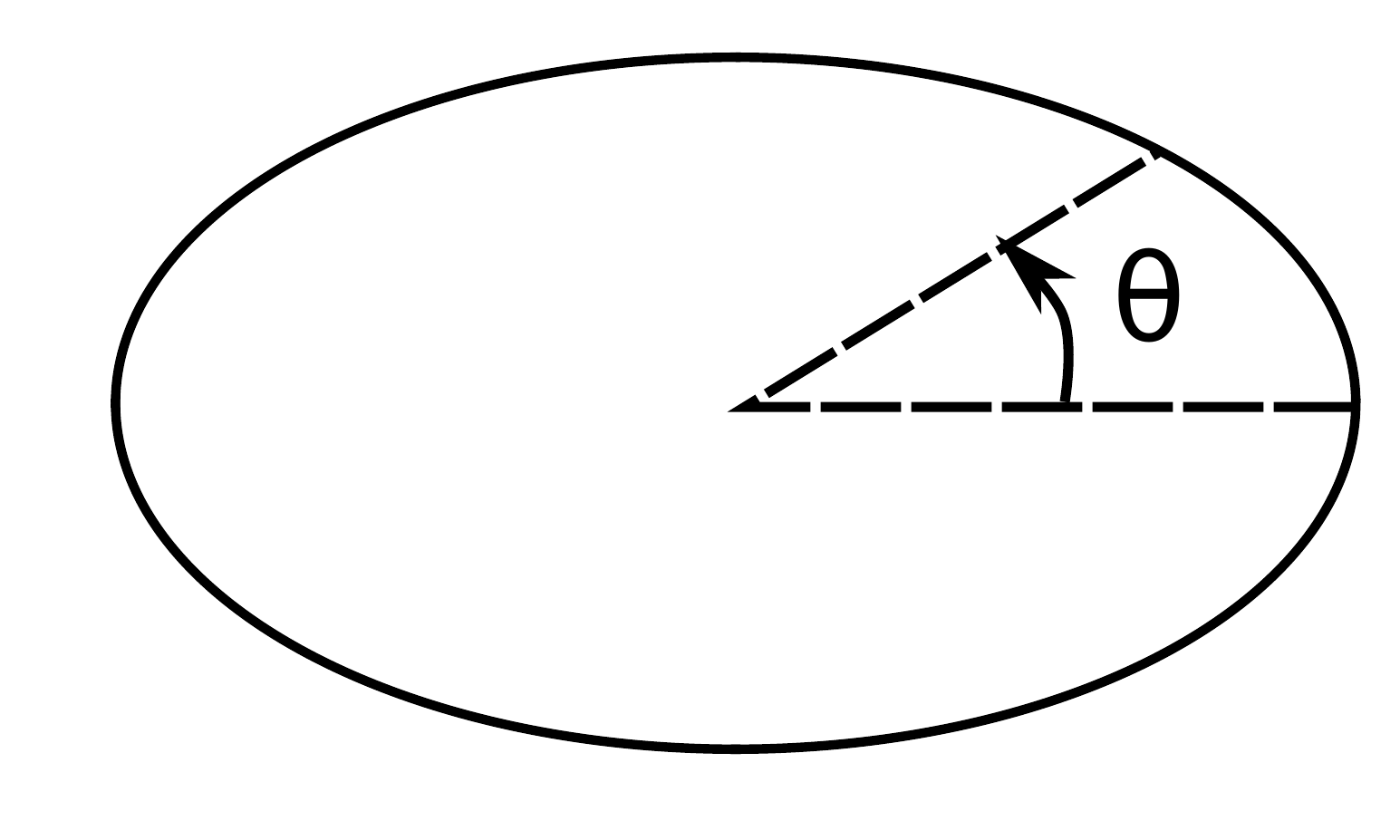}
        \hfill
        \caption{}\label{fig:curvature1}
    \end{subfigure}
    \begin{subfigure}[b]{0.45\linewidth}
        \includegraphics[width=\linewidth]{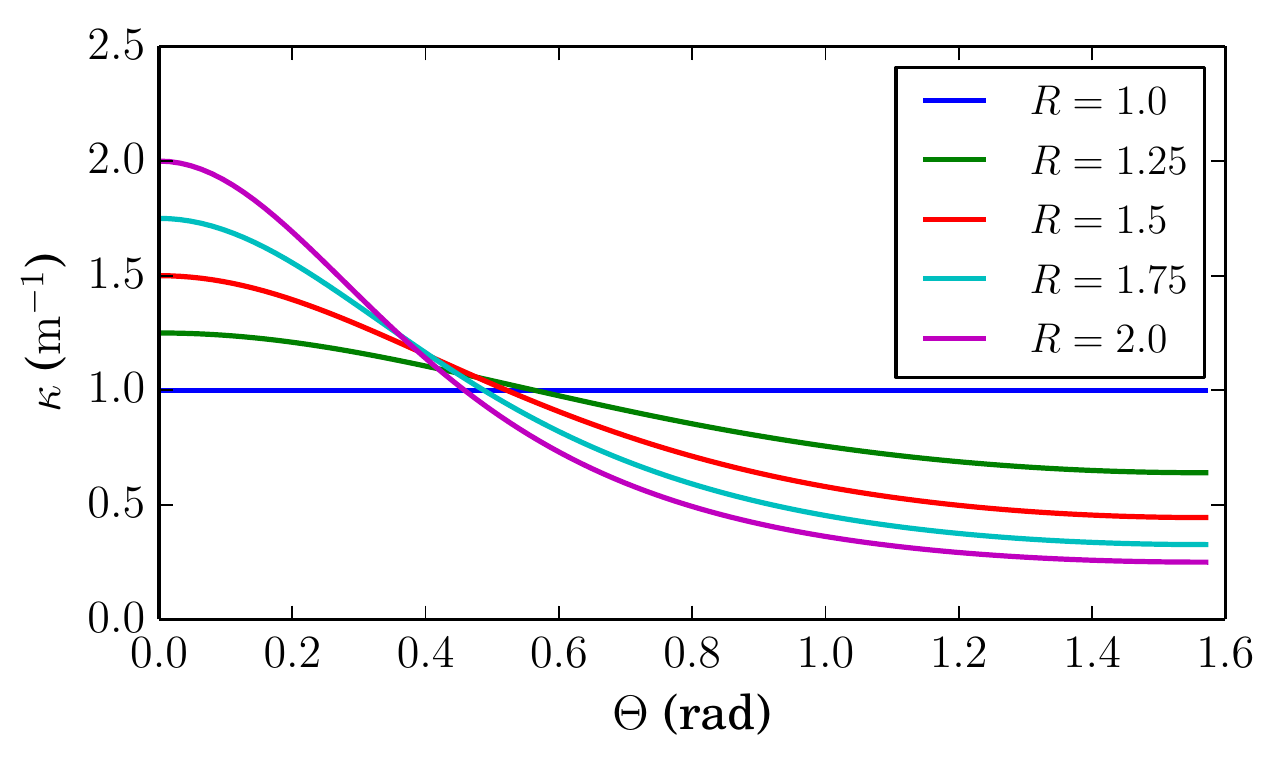}
        \hfill
        \caption{}\label{fig:curvature2}
    \end{subfigure}
    \caption{(a) Schematic of an ellipse in polar coordinates; (b) (color online) plot of the mean curvature $\kappa$ of the ellipse boundary versus polar angle $\theta$. \label{fig:curvature}}
\end{figure}

Finally, the fully nematic domain relaxes towards the equilibrium state, shown in the final sets of images in Fig.~\ref{fig:formation} for the domains with aspect ratios $R=\{1.05,1.6, 2\}$.
This relaxation involves simultaneous motion of the disclinations along the major axis and bulk reorientation.
The relaxation mechanism for circular domains has been shown to be governed by the competition of bulk elasticity and surface anchoring strength \cite{Sharma2003}.
Elliptic domains introduce an additional contribution: the variation of the curvature of the boundary.
Fig. \ref{fig:equilibrium_distance} shows the evolution of the distance between defect cores for the formation process of each of the domains simulated.
At equilibrium, the defect separation distance is found to increase with increasing aspect ratio which occurs without any changes in bulk elasticity or surface anchoring strength.
This behavior can be explained by quantifying the mean curvature imposed by the elliptic boundary conditions.
Figs. \ref{fig:curvature1}-\ref{fig:curvature2} show the schematic of an ellipse in polar coordinates and the mean curvature $\kappa$ as a function of $\theta$ for ellipses of increasing aspect ratio.
As aspect ratio increases, the curvature of the boundary regions increasingly becomes distinct: (i) a high-curvature region outward by the ellipse focal points and (ii) low-curvature regions elsewhere.
Thus as aspect ratio increases the combination of geometry and anchoring effects increasingly impose highly localized deformation of the nematic in the focal regions of the ellipse, which results in equilibrium textures with defects located in these regions.

\subsection{Field-switching Dynamics}

Field-switching dynamics simulations were performed for the range of aspect ratios $R=(1,2]$ using the results from Sec. \ref{sec:formation} as initial conditions.
Electric field orientation was chosen to be parallel to the major axis which results in the most significant field-driven effect on domain texture.
The results of these simulations are presented and discussed by first focusing on the equilibrium domain textures during application of the electric field and after its release.
Following this, the dynamics of the transition from the initial equilibrium domain to the field-driven domain and after release of the field are presented and discussed.
In order to interpret the complex textures and texture dynamics resulting from these transient simulations, the \emph{droplet order parameter} is introduced \cite{Kelly1994},
\begin{equation}
    \bm{Q}_{d} = V^{-1}\int_{V} \bm{Q} dV 
\end{equation}
which can be further decomposed into the \emph{droplet director} $\bm{n}_{d}$ and \emph{droplet scalar order parameter} $S_{d}$.
In all elliptic cases the droplet director $\bm{n}_{d}$ is initially parallel to the minor axis due to the homeotropic anchoring conditions.
The magnitude of $S_{d}$ relates to the optical properties such that as $S_{d}$ decreases, the domain more efficiently scatters light and as it increases, the domain more efficiently transmits light.

\subsubsection{Field-driven Equilibrium Textures}

\begin{figure}[h!]
    \begin{subfigure}[b]{0.2065\linewidth}
        \includegraphics[width=\linewidth]{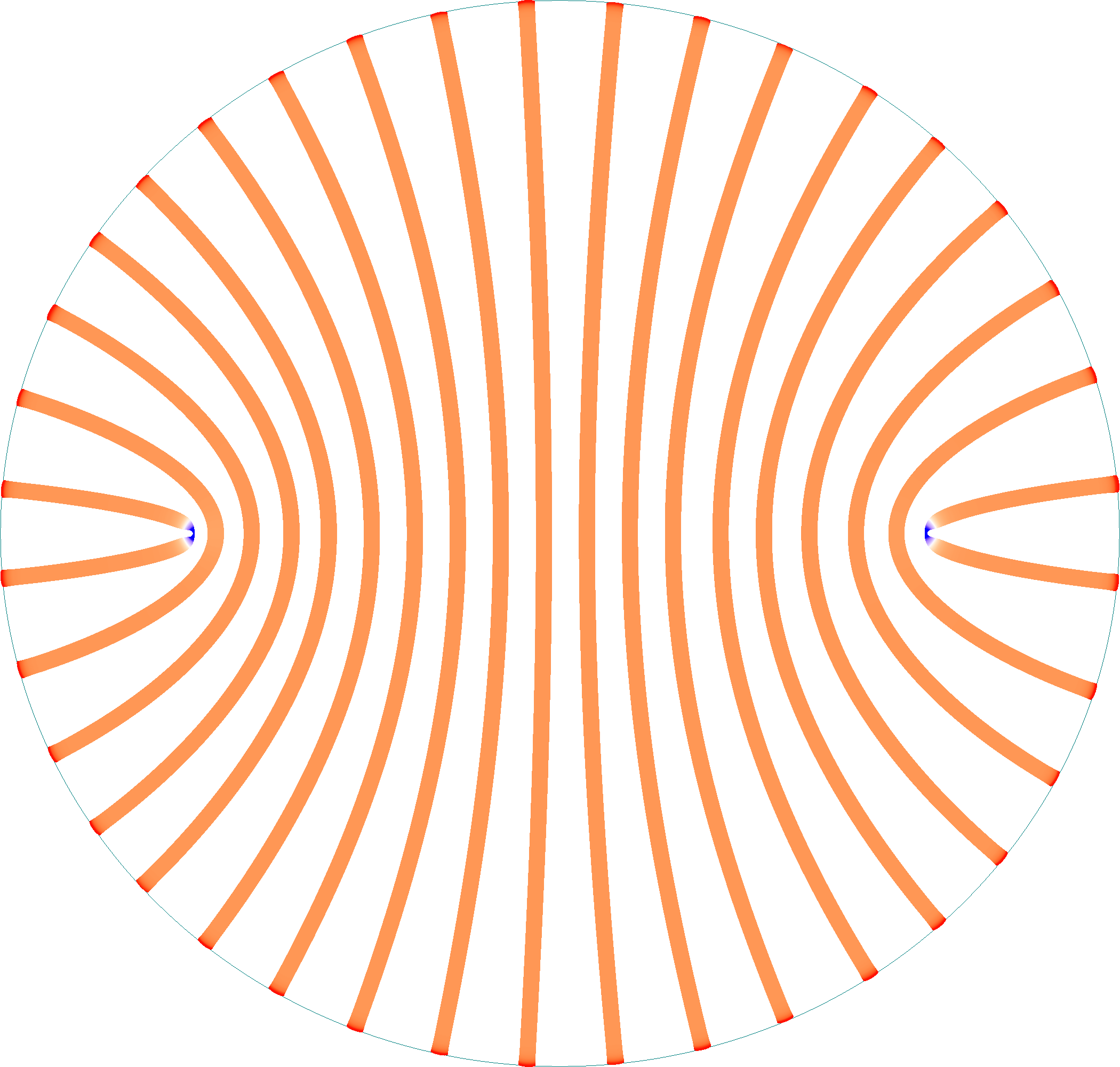}
        \caption{}
    \end{subfigure}
    \begin{subfigure}[b]{0.3304\linewidth}        
        \includegraphics[width=\linewidth]{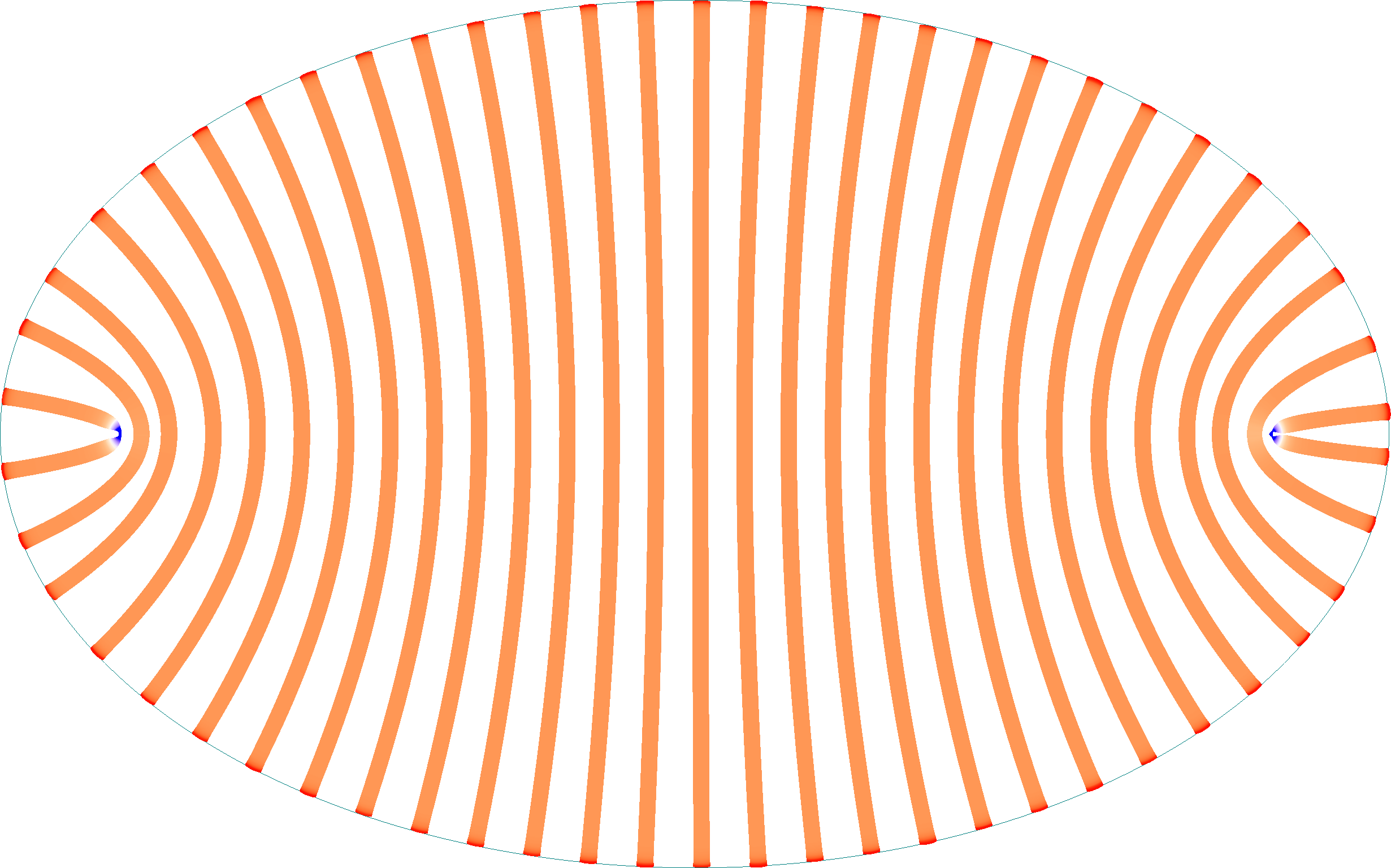}
        \caption{}
    \end{subfigure}
    \begin{subfigure}[b]{0.413\linewidth}    
        \includegraphics[width=\linewidth]{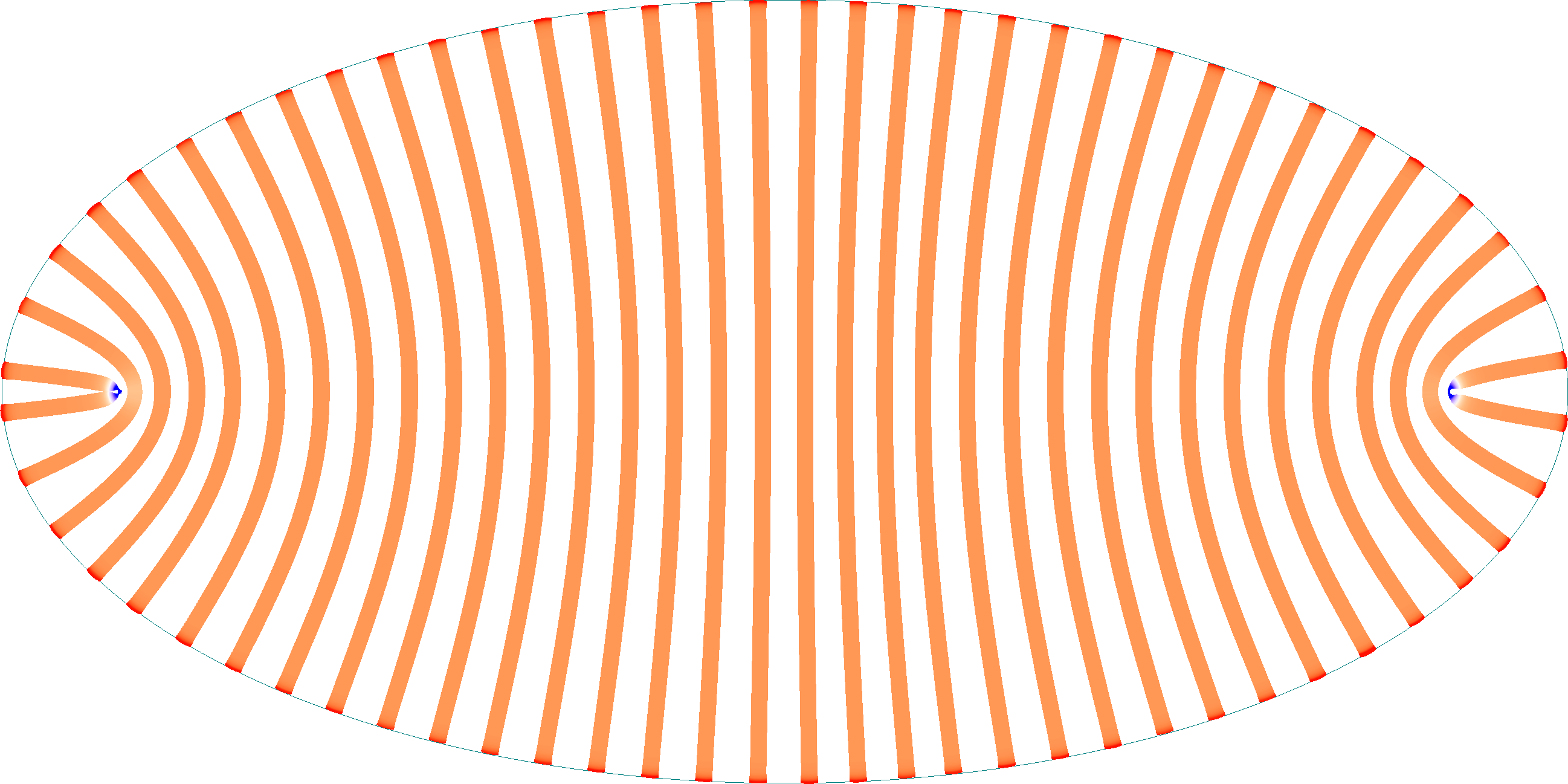}
        \caption{}
    \end{subfigure}
    \includegraphics[scale=0.07]{scalebar_horiz}
    \caption{(Color online) Hyperstreamline visualizations of field-driven textures with $E < E_c$ for domains with aspect ratio (a) $R = 1.05$, (b) $R = 1.6$, and (c) $R = 2.0$. \label{fig:driven_below}}
\end{figure}

\begin{figure}[h!]
    \begin{subfigure}[b]{0.2065\linewidth}
        \includegraphics[width=\linewidth]{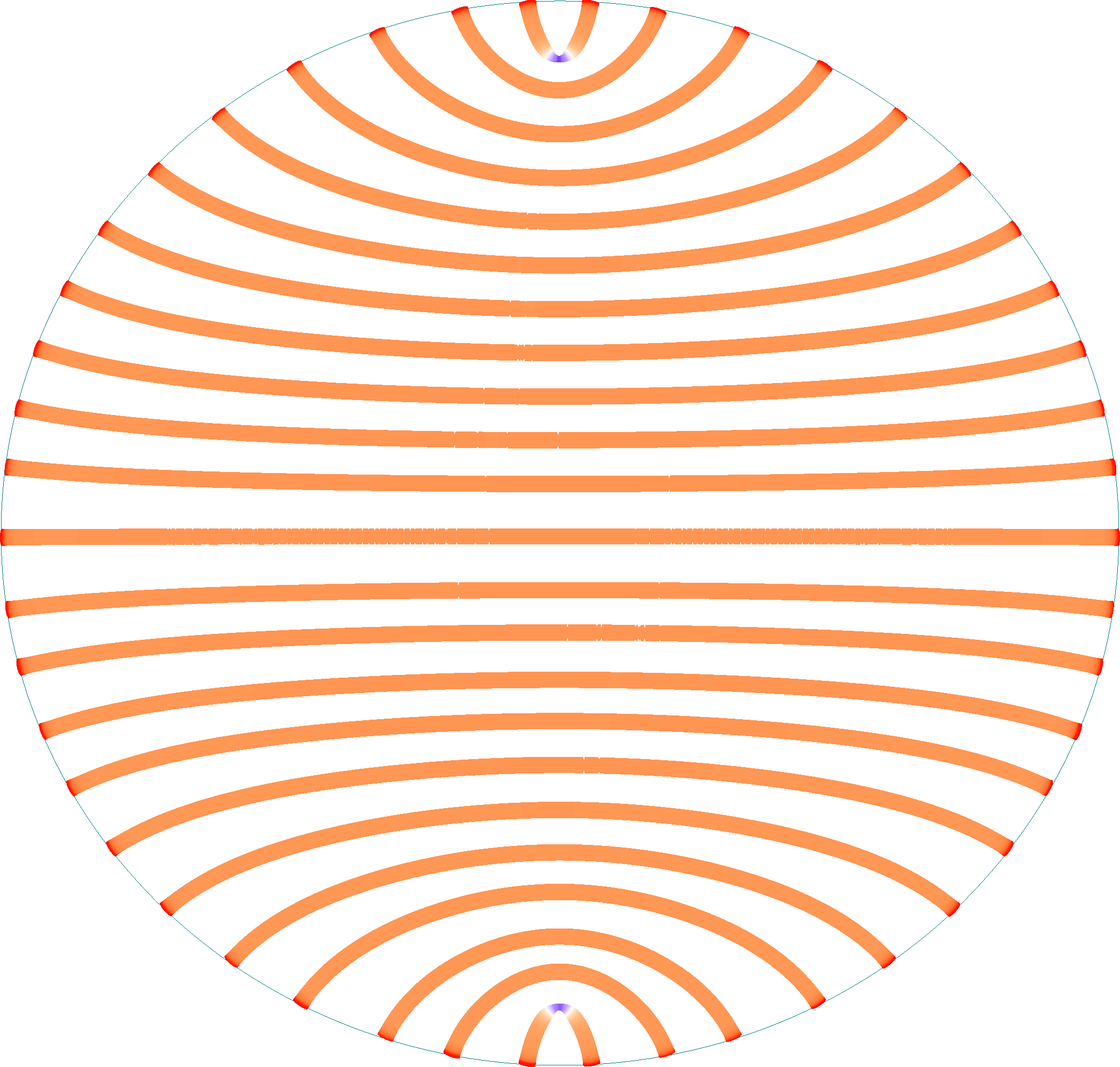}
        \caption{}
    \end{subfigure}
    \begin{subfigure}[b]{0.3304\linewidth}        
        \includegraphics[width=\linewidth]{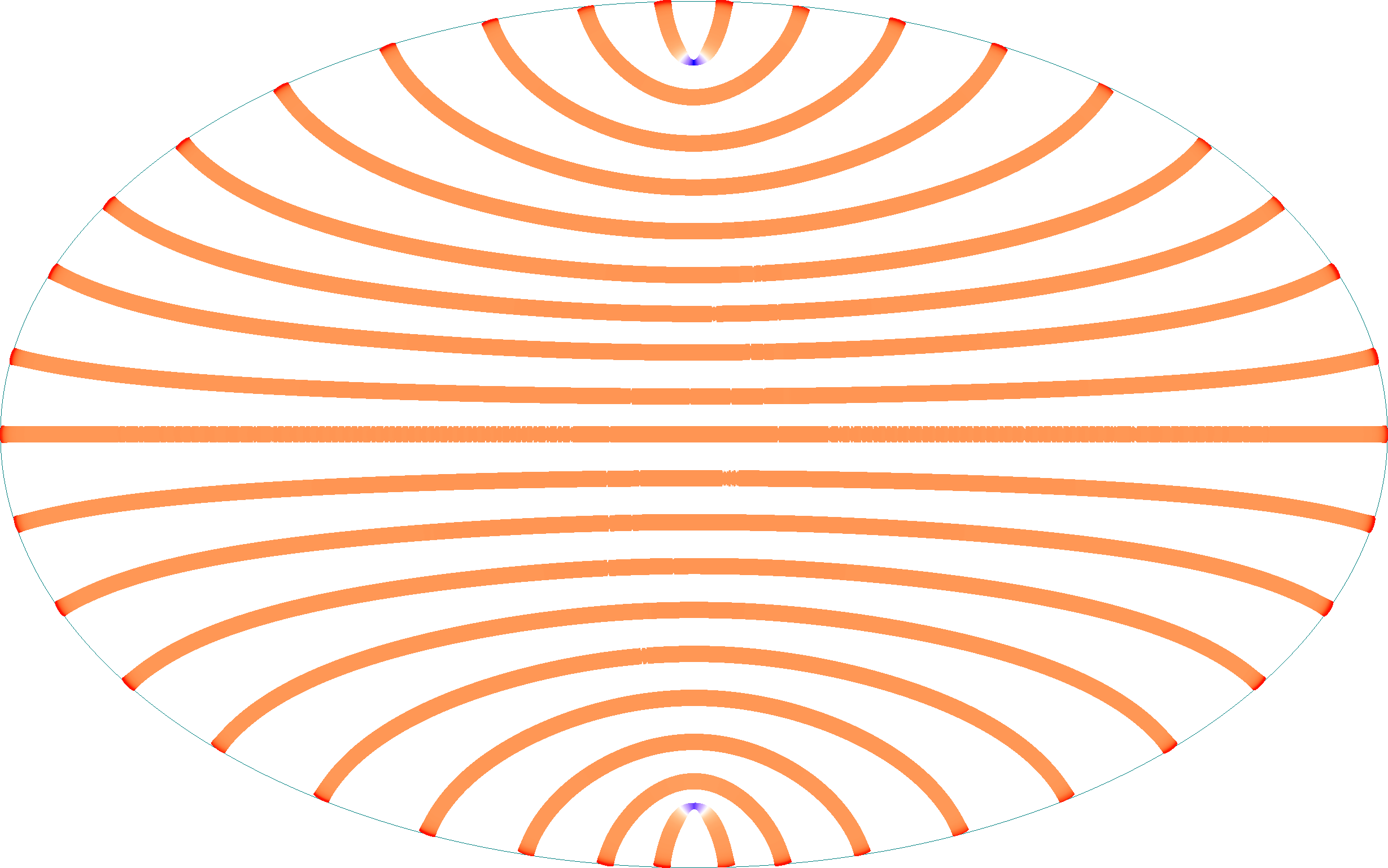}
        \caption{}
    \end{subfigure}
    \begin{subfigure}[b]{0.413\linewidth}    
        \includegraphics[width=\linewidth]{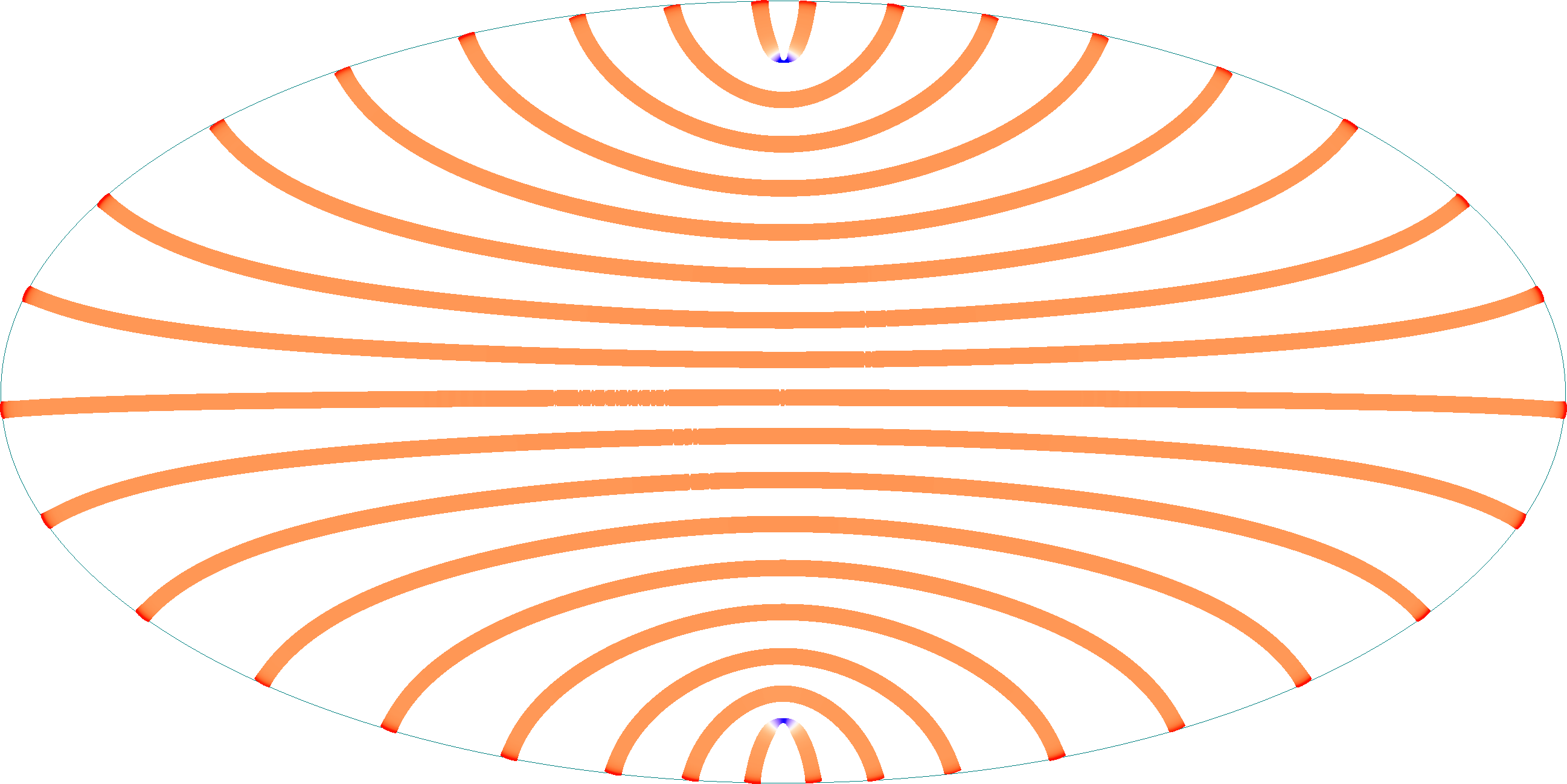}
        \caption{}
    \end{subfigure}
    \includegraphics[scale=0.07]{scalebar_horiz}
    \caption{(Color online) Hyperstreamline visualizations of field-driven textures with $E > E_c$ for domains with aspect ratio (a) $R = 1.05$, (b) $R = 1.6$, and (c) $R = 2.0$. \label{fig:driven_above}}
\end{figure}

Simulations of application and release of an electric field, with strengths ranging from $(0 \si{\volt\per\micro\meter}, 5 \si{\volt\per\micro\meter}] $ were performed.
Two different field-switching regimes were observed where the droplet director $\bm{n}_{d}$ either remained constant (low field strength) or reoriented (high field strength).
Visualizations of the equilibrium Q-tensor fields for both regimes are shown in Fig. \ref{fig:driven_below} and  Fig. \ref{fig:driven_above}, respectively.
These results support the assumption that a critical field strength $E_{c}$ exists depending on the domain aspect ratio, anchoring strength and LC material properties.

For the $E < E_{c}$ regime in the field-driven state, there is a small response in the domain texture resulting in a change in the droplet scalar order parameter $S_{d}$, but not the droplet director.
Defects are driven inwards along the major axis which results in larger field-aligned focal regions; this mechanism similar to the domain texture prior to the bulk relaxation regime observed during formation.
Upon release of the field the domain is restored to the initial equilibrium texture resulting from the formation process.

For the $E \ge E_{c}$ regime in the field-driven state, there is a large response of the domain where disclinations transition to be aligned along the minor axis and the domain becomes strongly field-aligned through reorientation of $\bm{n}_{d}$.
This is achieved through the domination of electric field forces over surface anchoring forces, although defect ``escape'' is not observed as it could be for very strong electric field or very weak surface anchoring conditions \cite{Crawford1992}.
Upon release of the field the domain is restored to the initial equilibrium texture resulting from the formation process.

\begin{figure}[h!]      
    \includegraphics[width=0.5\linewidth]{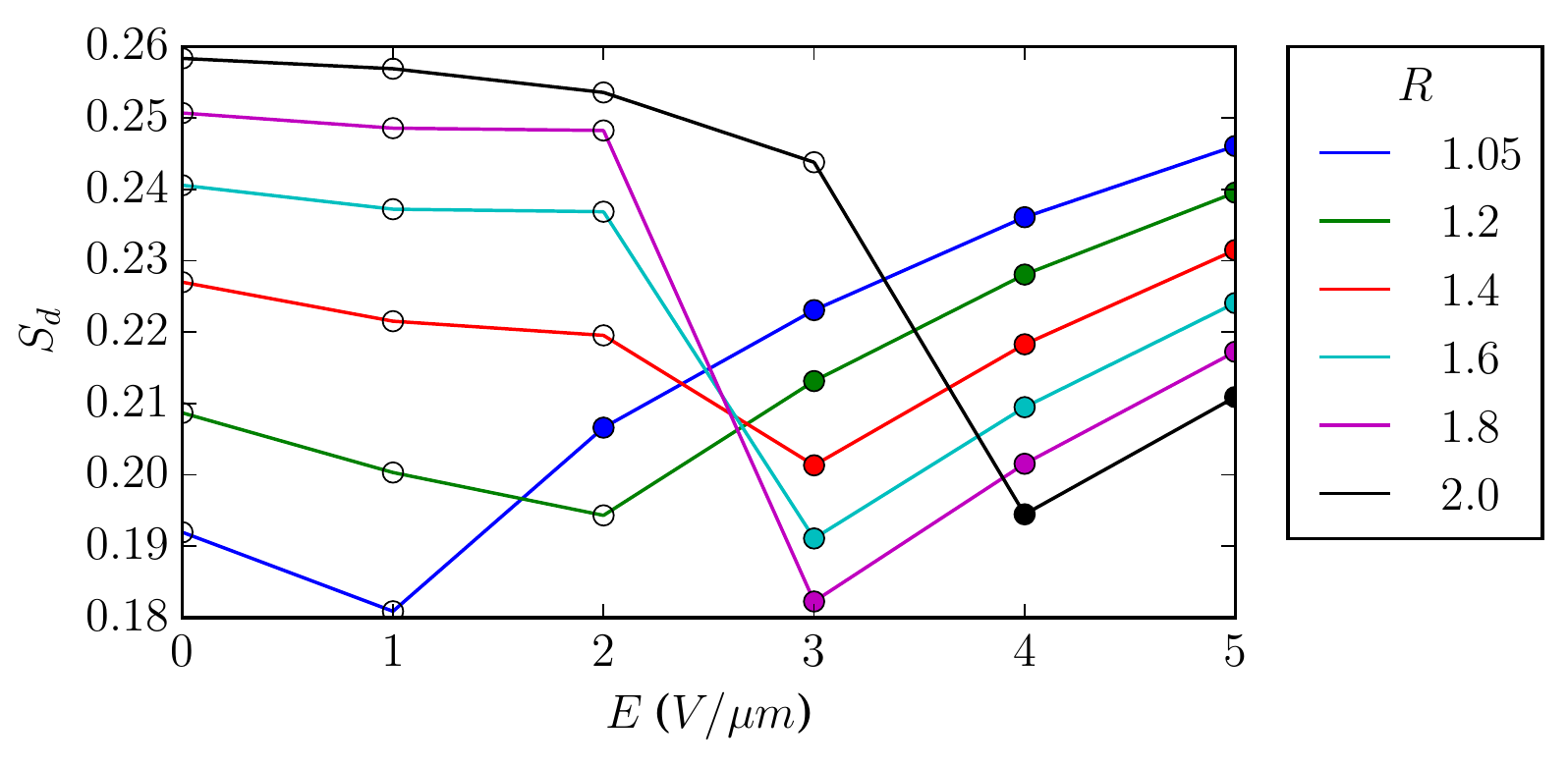}
    \caption{(Color online) Plot of droplet scalar order parameter $S_{d}$ versus electric field strength $E$ for $R=(1,2]$ where $\circ$ and $\bullet$ glyphs correspond to $\bm{n}_{d} \perp \bm{E}$ and $\bm{n}_{d} \parallel \bm{E}$, respectively. \label{fig:equilibrium_order_parameter}}
\end{figure}

Fig. \ref{fig:equilibrium_order_parameter} shows droplet scalar order parameters $S_{d}$ for simulated equilibrium domains at different field strengths both below and above $E_{c}$. 
As electric field strength is increased (but still held below $E_{c}$) the effect is minimal on $S_{d}$.
This can be explained through observation of the field-driven domain textures in Fig. \ref{fig:driven_below}; while the area aligned with the field (focal regions) increases in size and uniformity, this is achieved through a simultaneous reduction in the central region that is aligned orthogonal to the field.
Thus the net increase in nematic order in the focal regions is almost completely offset by decreased order in the central region.

As electric field strength is increased above $E_{c}$ there is a significant deformation of the domain texture, both with respect to $S_{d}$ and $\bm{n}_{d}$.
This response is similar to the Fredericks transition observed in planar LC domains exposed to an external field \cite{deGennes1995}.
In all simulations the droplet director $\bm{n}_{d}$ is observed to reorient parallel to the applied field vector (major axis).
The observed response of $S_{d}$ after application of the field is more complex.
For low aspect ratios $R < 1.5$ the electric field is found to result in a field-aligned nematic domain with droplet scalar order parameter greater than in the absence of the field.
This is the traditional mechanism associated with PDLC films used for privacy glass, where in the ``off'' state they scatter substantially more than in the ``on'' state.
For higher aspect ratios $R > 1.5$ the \emph{opposite} behavior is observed; even though the domain texture is field-aligned, the droplet scalar order parameter is substantially lower in the field-driven state compared to without the field.
This corresponds to an increase in light scattering, regardless of the orientation of the droplet director $\bm{n}_{d}$.
This ``reverse-mode''-like mechanism is typically associated with PDLC films formed using a nematic LC with negative dielectric anisotropy \cite{Coates1993}.

\subsubsection{Field-driven Dynamics Textures}

\begin{figure}[h!]
    \begin{subfigure}[b]{0.31\linewidth}
        \includegraphics[width=\linewidth]{ar2_equilibrium}
        \caption{}
    \end{subfigure}
    \begin{subfigure}[b]{0.31\linewidth}        
        \includegraphics[width=\linewidth]{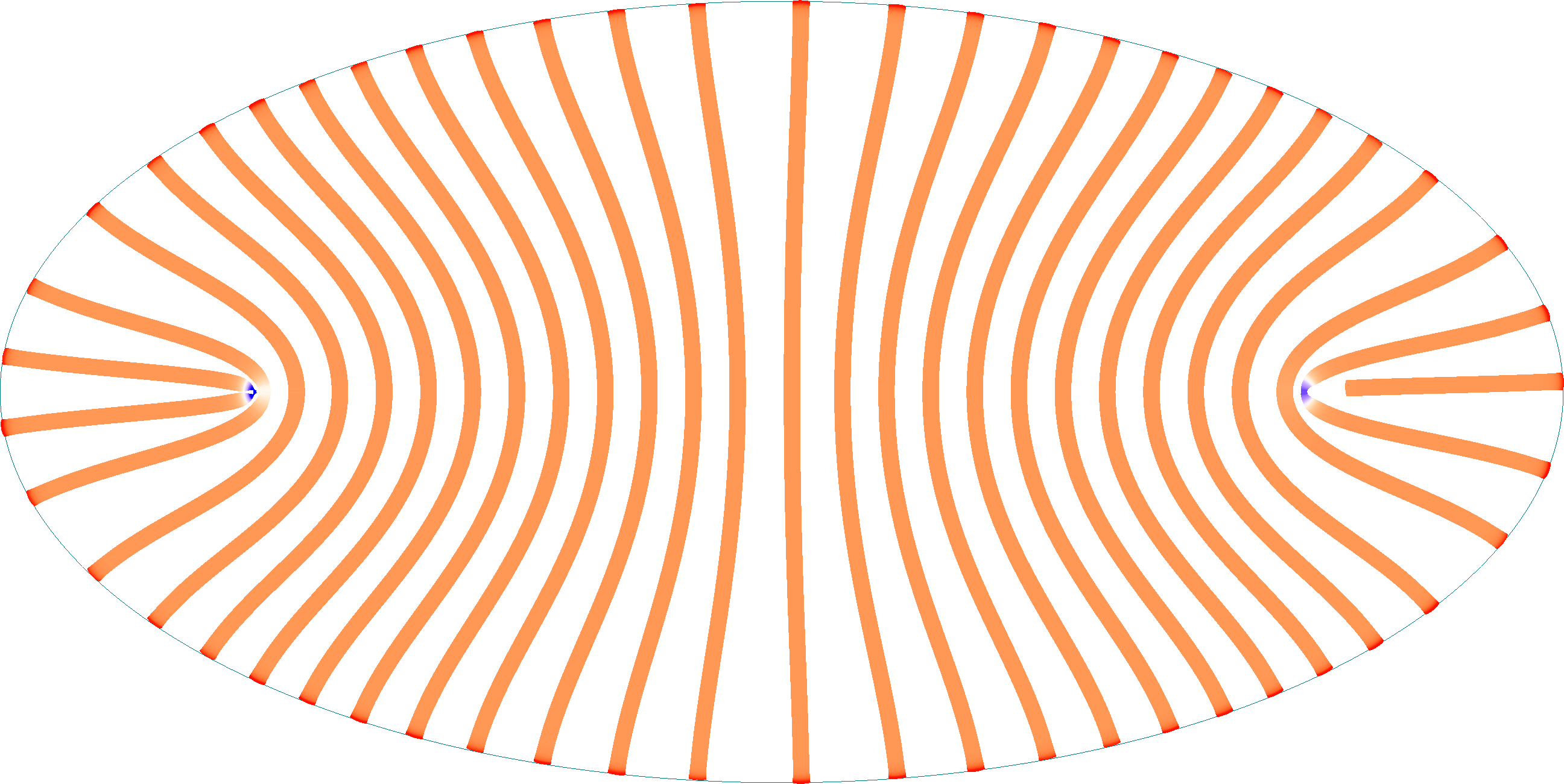}
        \caption{}
    \end{subfigure}
    \begin{subfigure}[b]{0.31\linewidth}    
        \includegraphics[width=\linewidth]{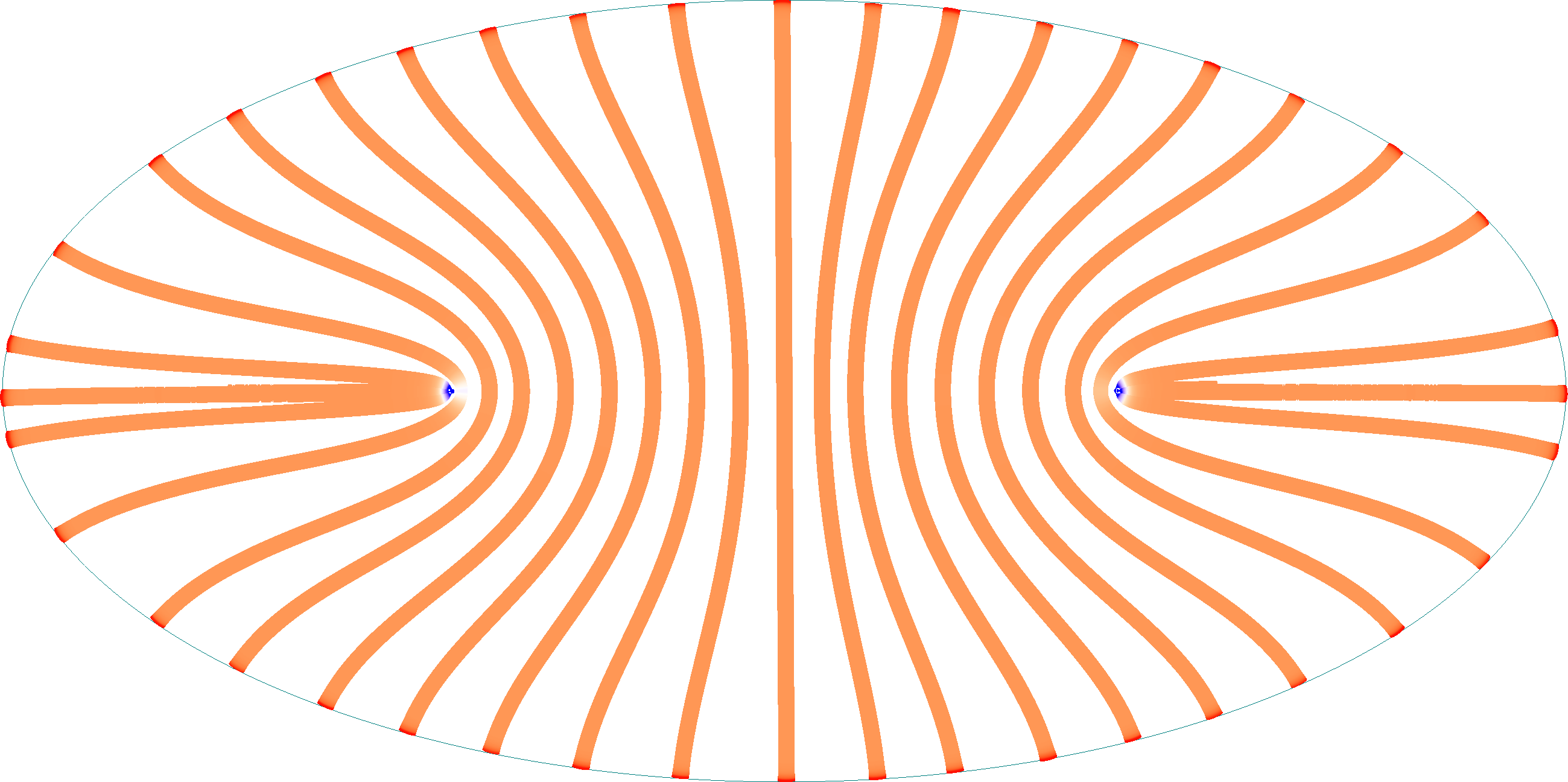}
        \caption{}
    \end{subfigure}\\
    \begin{subfigure}[b]{0.31\linewidth}
        \includegraphics[width=\linewidth]{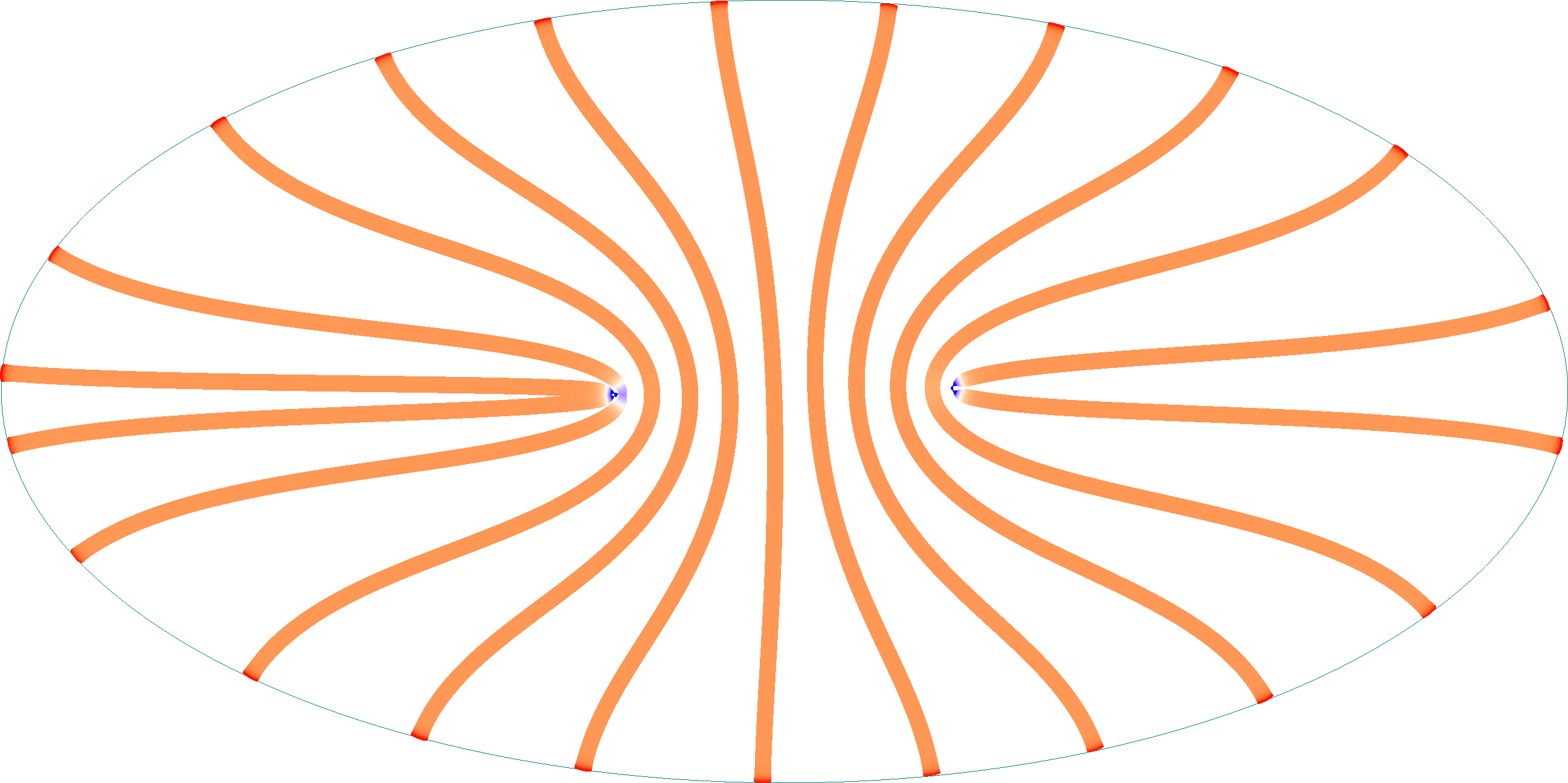}
        \caption{}
    \end{subfigure}
    \begin{subfigure}[b]{0.31\linewidth}        
        \includegraphics[width=\linewidth]{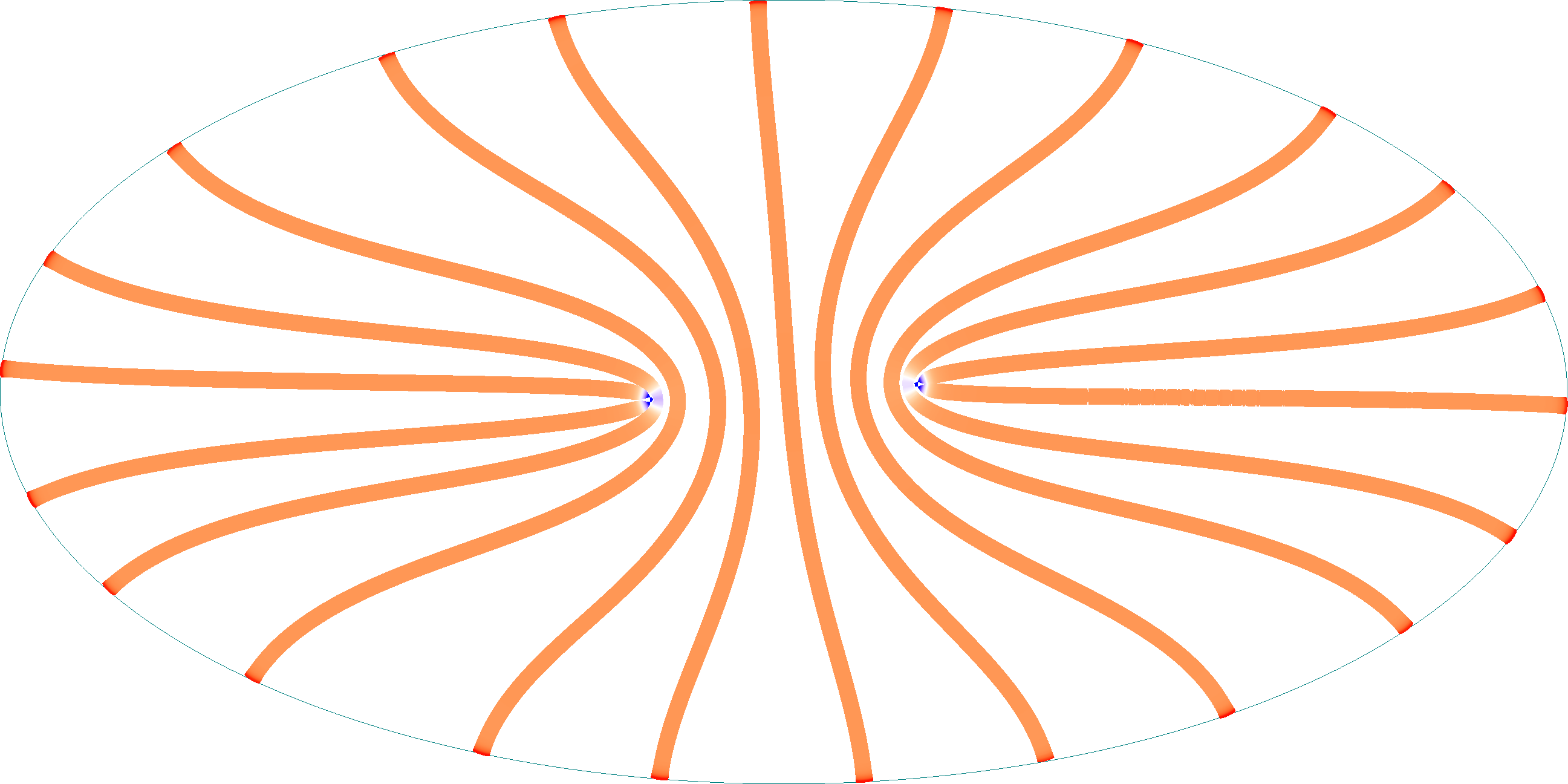}
        \caption{}
    \end{subfigure}
    \begin{subfigure}[b]{0.31\linewidth}    
        \includegraphics[width=\linewidth]{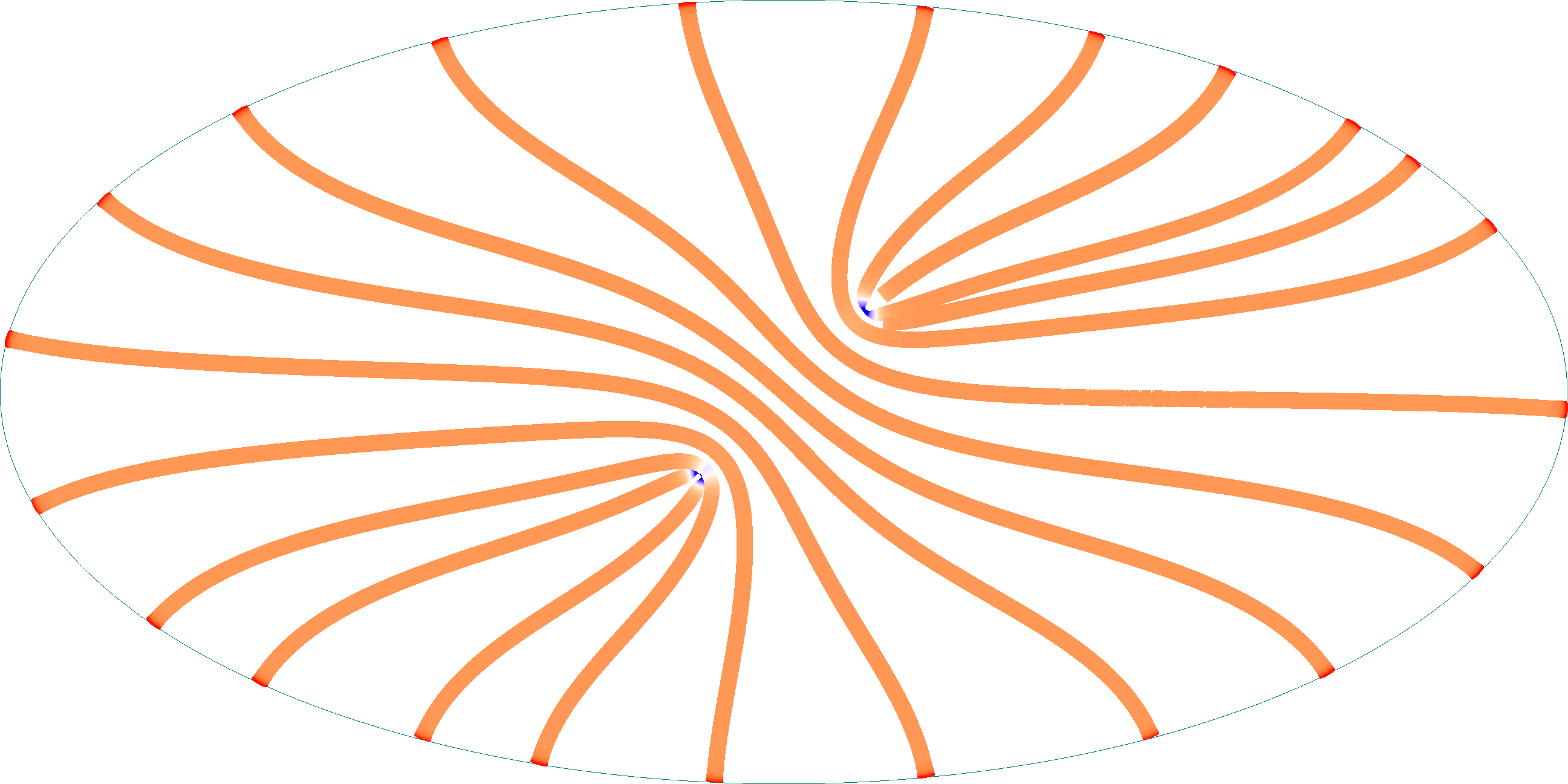}
        \caption{}
    \end{subfigure}\\
    \begin{subfigure}[b]{0.31\linewidth}    
        \includegraphics[width=\linewidth]{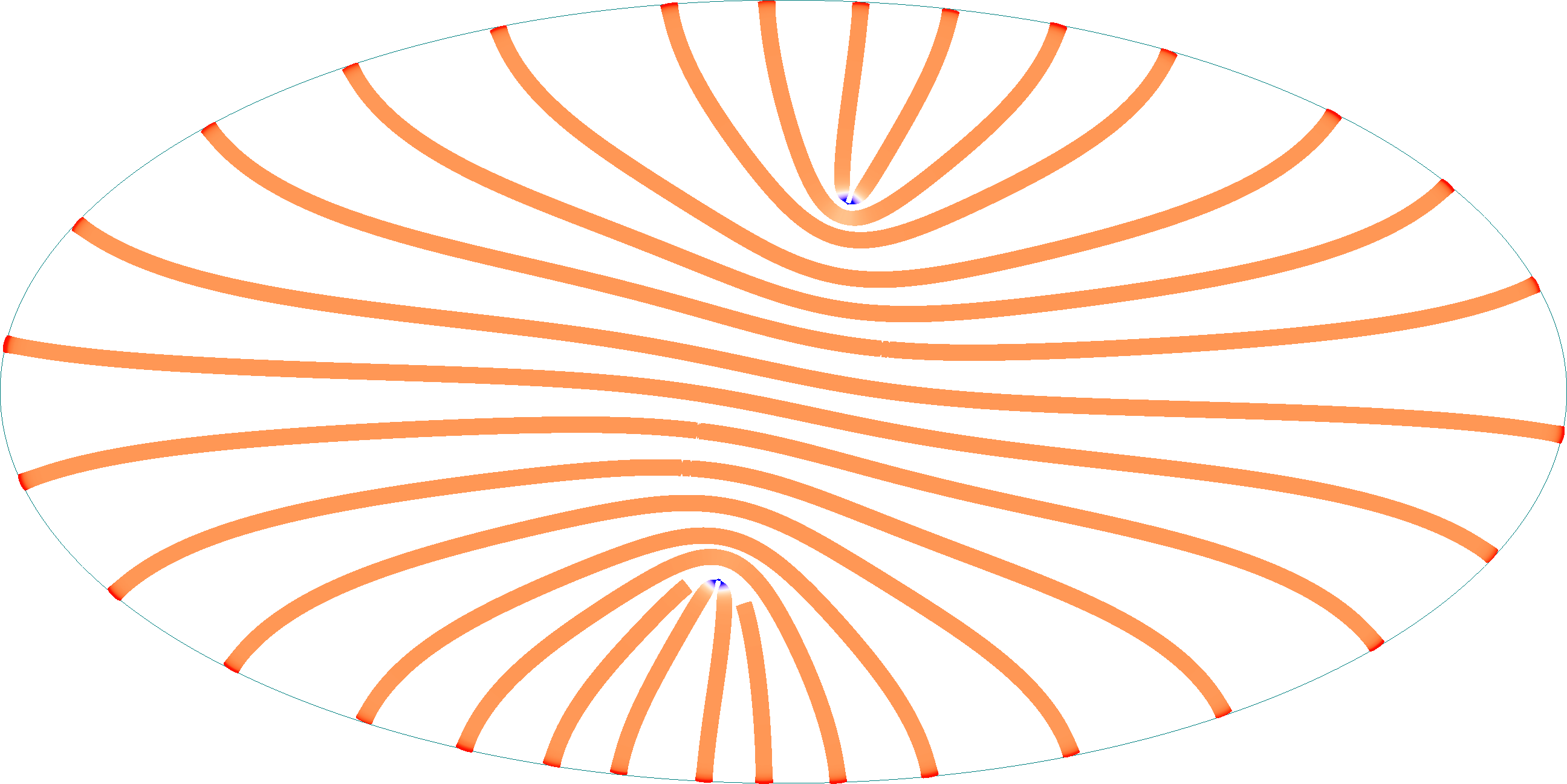}
        \caption{}
    \end{subfigure}
    \begin{subfigure}[b]{0.31\linewidth}    
        \includegraphics[width=\linewidth]{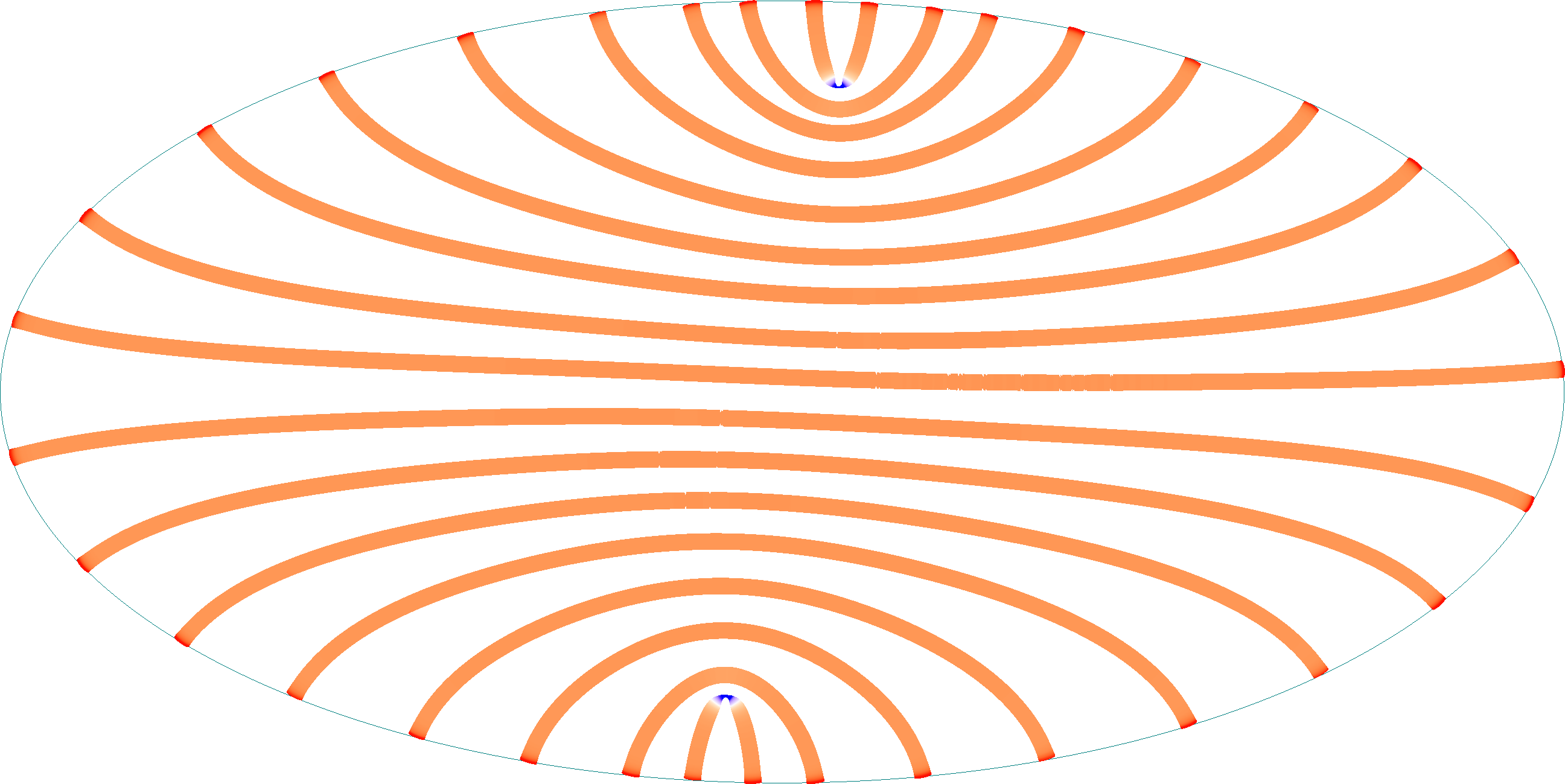}
        \caption{}
    \end{subfigure}
    \begin{subfigure}[b]{0.31\linewidth}    
        \includegraphics[width=\linewidth]{ar2_driven_equilibrium}
        \caption{}
    \end{subfigure}\\
    \includegraphics[scale=0.07]{scalebar_horiz}
    \caption{(Color online) Hyperstreamline visualizations of the field-on texture dynamics for the $R=2$ domain with $E = \SI{5}{V/\micro\metre}$ and $t = \{\SI{0}{}$, $\SI{5.25e5}{}$, $\SI{1.05e6}{}$, $\SI{1.55e6}{}$, $\SI{1.89e6}{}$, $\SI{2.11e6}{}$, $\SI{2.33e6}{}$, $\SI{2.71e6}{}$, $\ge \SI{5.00e6}{}\}$.\label{fig:driven_dynamics}}
\end{figure}

Fig. \ref{fig:driven_dynamics} shows simulation results of the field-driven dynamics for the $R = 2$ case for $E = 5 \si{\volt\per\micro\meter} > E_{c}$.
These results are representative of simulation results for all domains $R=(1,2]$.
A sequence of three distinct dynamic regimes were observed: bulk growth/recession, disclination/bulk rotation, and bulk relaxation.

\begin{figure}[h!]
    \centering
    \begin{subfigure}[b]{0.45\linewidth}
        \includegraphics[width=\linewidth]{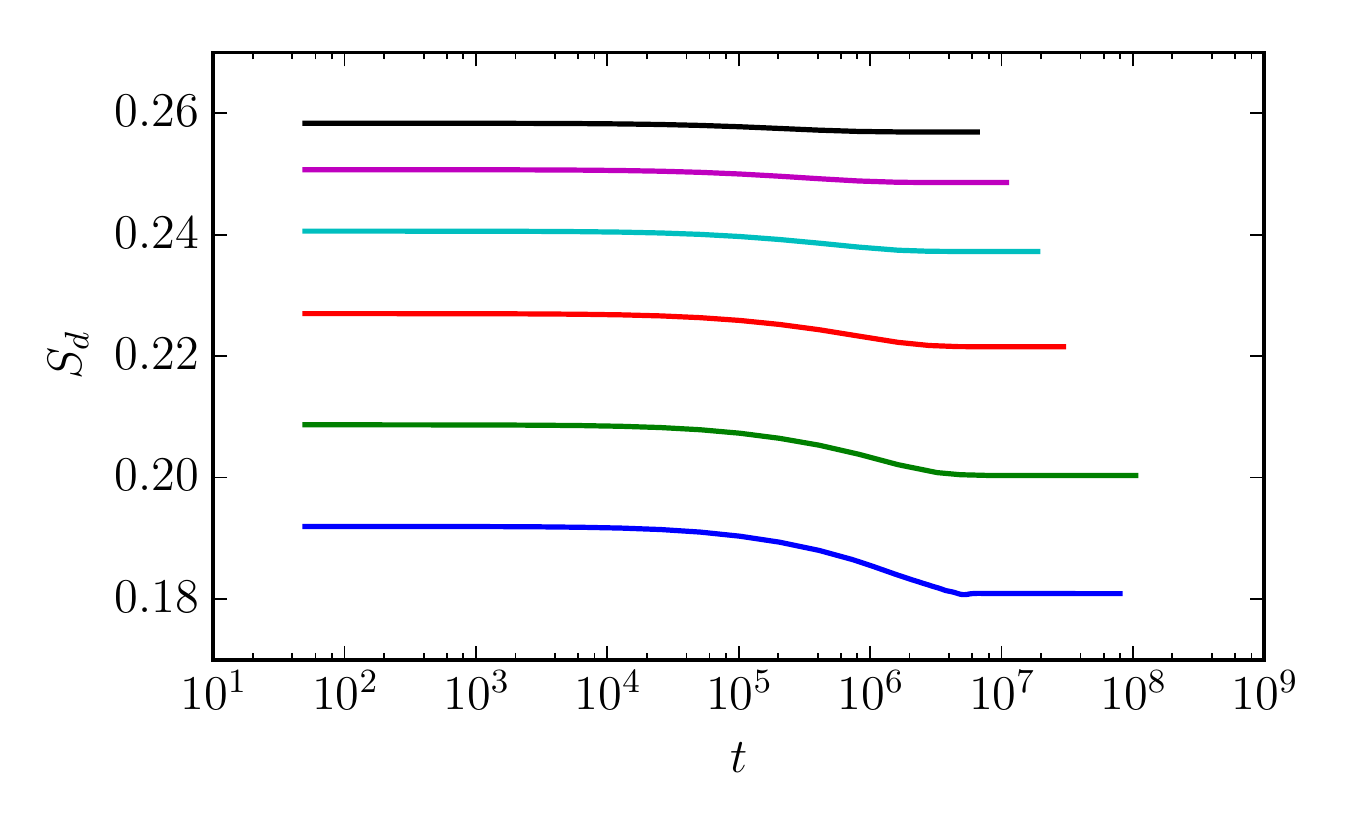}
        \caption{}\label{fig:below_driven}
    \end{subfigure}        
    \begin{subfigure}[b]{0.45\linewidth}
        \includegraphics[width=\linewidth]{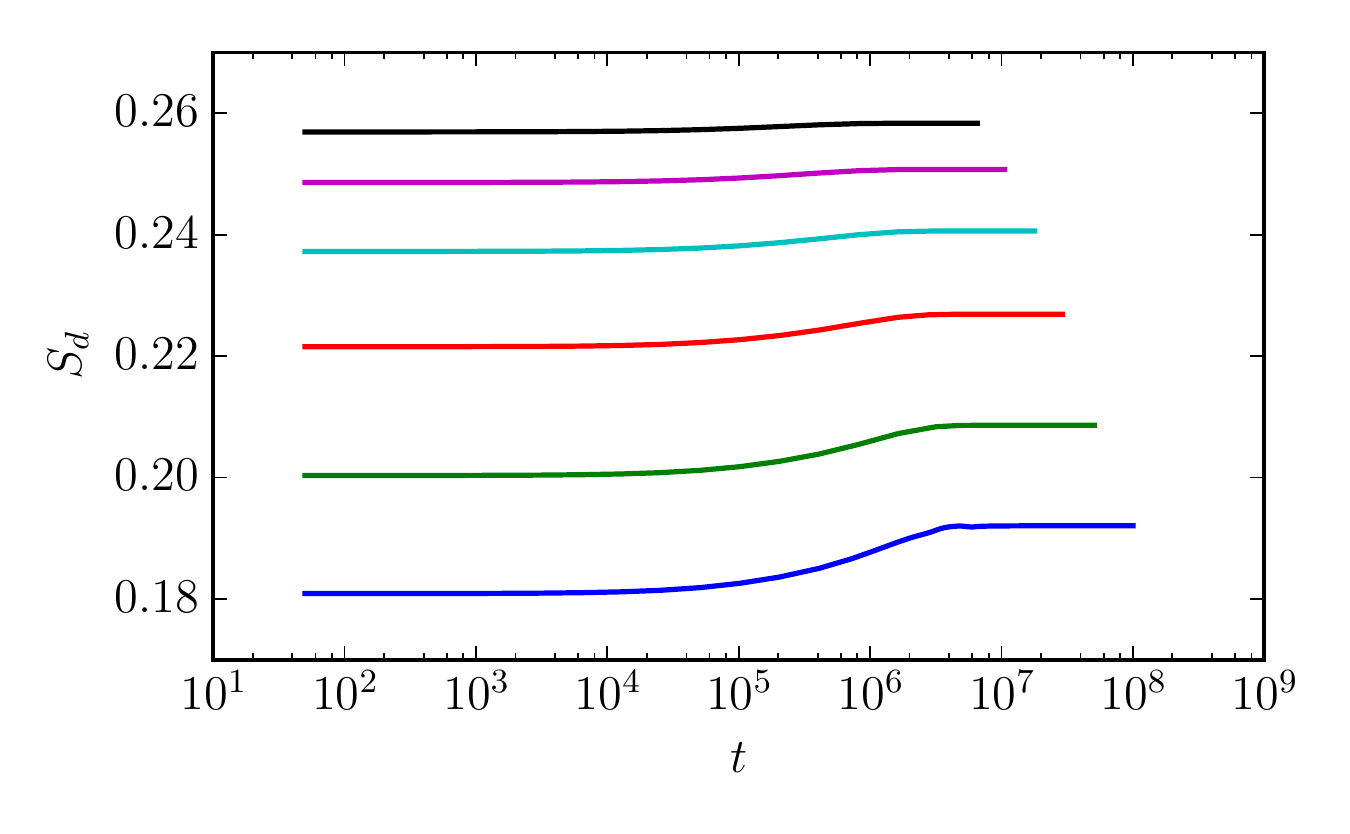}
        \caption{}\label{fig:below_release}
    \end{subfigure}\\
    \begin{subfigure}[b]{0.45\linewidth}
        \centering
        \includegraphics[width=\linewidth]{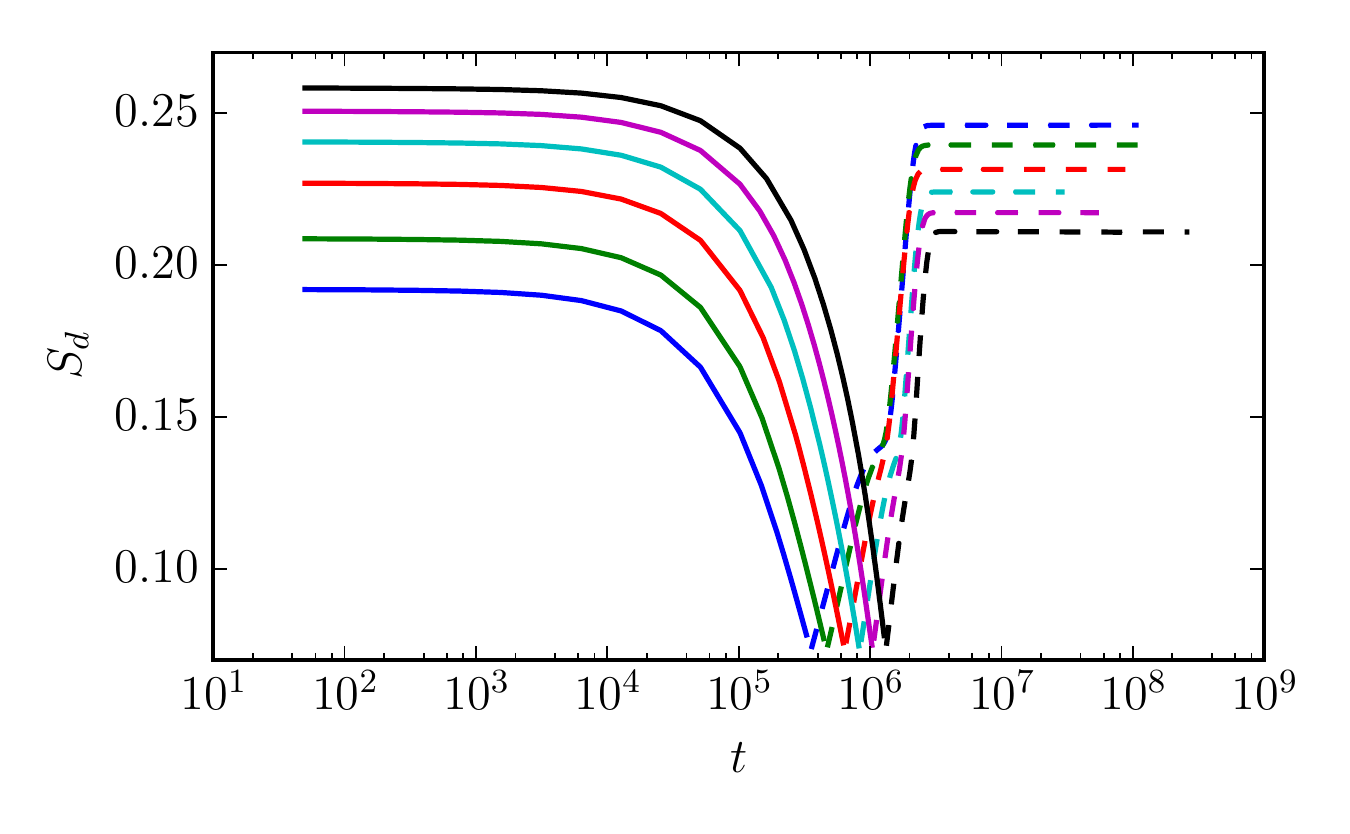}
        \caption{}\label{fig:above_driven}
    \end{subfigure}        
    \begin{subfigure}[b]{0.45\linewidth}
        \includegraphics[width=\linewidth]{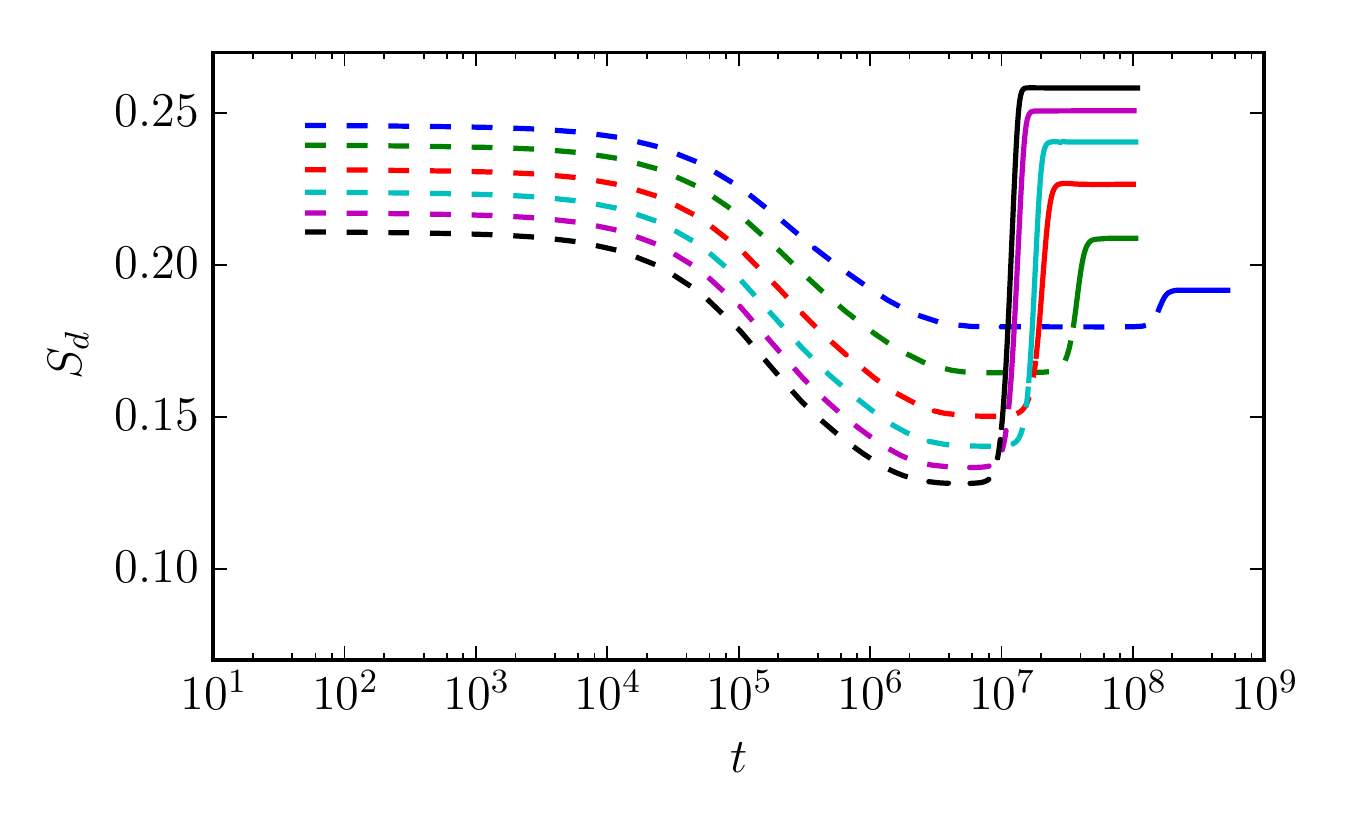}
        \caption{}\label{fig:above_release}
    \end{subfigure}\\
    \includegraphics[width=0.6\linewidth]{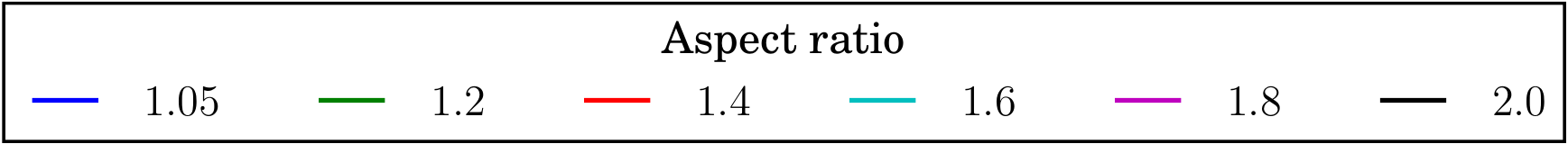}
    \caption{(Color online) Droplet scalar order parameter evolution versus time for domains with a range of aspect ratios $R=(1,2]$ with electric field applied (a,c) and following release (b,d): (a) $E = \SI{1}{V/\micro\metre} < E_c$ (driven); (b) $E = \SI{1}{V/\micro\metre} < E_c$ (release); (c) $E = \SI{5}{V/\micro\metre} > E_{c}$ (driven); and (d) $E = \SI{5}{V/\micro\metre} > E_{c}$ (release). Droplet director orientation $\bm{n}_{d}$ is indicated by line type: solid and dashed correspond to $\bm{n}_{d} \perp \bm{E}$ and $\bm{n}_{d} \parallel \bm{E}$, respectively. \label{fig:dynamic_order_parameter}}
\end{figure}

The \emph{bulk growth/recession regime} involves the simultaneous growth of the field-aligned focal regions and recession of the field-orthogonal central region which involves the motion of the disclinations inwards along the major axis.
This regime was observed for both the $E < E_{c}$ and $E > E_{c}$ cases, where in the former the net change in the domain texture was minimal (see Fig. \ref{fig:equilibrium_order_parameter}). 
In the latter case ($E > E_{c}$) as shown in Fig. \ref{fig:above_driven}, this regime involves a monotonic decrease in the droplet scalar order parameter $S_{d}\rightarrow 0$, corresponding to a radial texture (maximal light scattering).

The \emph{disclination/bulk rotation} regime follows the bulk growth/recession regime; as the distance between disclination defects decreases (along the major axis) the repulsive nematic elastic forces approach that of the applied field.
At that point, the defect separation distance becomes constant and rotation occurs.
The initiation of this regime is found to involve a sharp transition of the droplet director from being orthogonal to parallel to the field direction (Fig. \ref{fig:above_driven}).
This discontinuous transition of $\bm{n}_{d}$ is enabled by the radial texture of the droplet, where $S_{d} = 0$ making $\bm{n}_{d}$ a degenerate quantity.
Disclinations and bulk nematic texture simultaneously rotate about the center of the domain, increasing domain field alignment rapidly (Fig. \ref{fig:above_driven}).

The \emph{bulk relaxation} regime involves simultaneous rotation and expulsion of the disclinations from the central region along the minor axis.
The rotation process ceases as disclinations approach the boundaries and surface anchoring forces balance out bulk elastic forces.
Higher electric field strengths likely exist which would overcome surface anchoring and result in an ``escape'' texture \cite{Crawford1992}, but they exceed the field strengths typically used in PDLC applications \cite{Bronnikov2013}, the focus of this study.

The observed mechanism predicted by simulations provides a more refined understanding of the mechanism predicted by Drzaic \cite{Drzaic1988}.
As opposed to short timescale bulk reorientation followed by long timescale defect motion, simulations predict that the mechanism is instead through the \emph{simultaneous} motion of disclinations, growth of field-aligned regions, and recession of nonaligned regions.
As originally predicted by Drzaic, the long timescale component of the mechanism is through the motion of defects transitioning from one axis of the domain to the other, but the short timescale mechanism also involves linear defect motion and reorientation dynamics in the outer regions of the domain.

\begin{figure}[h!]
    \begin{subfigure}[b]{0.31\linewidth}
        \includegraphics[width=\linewidth]{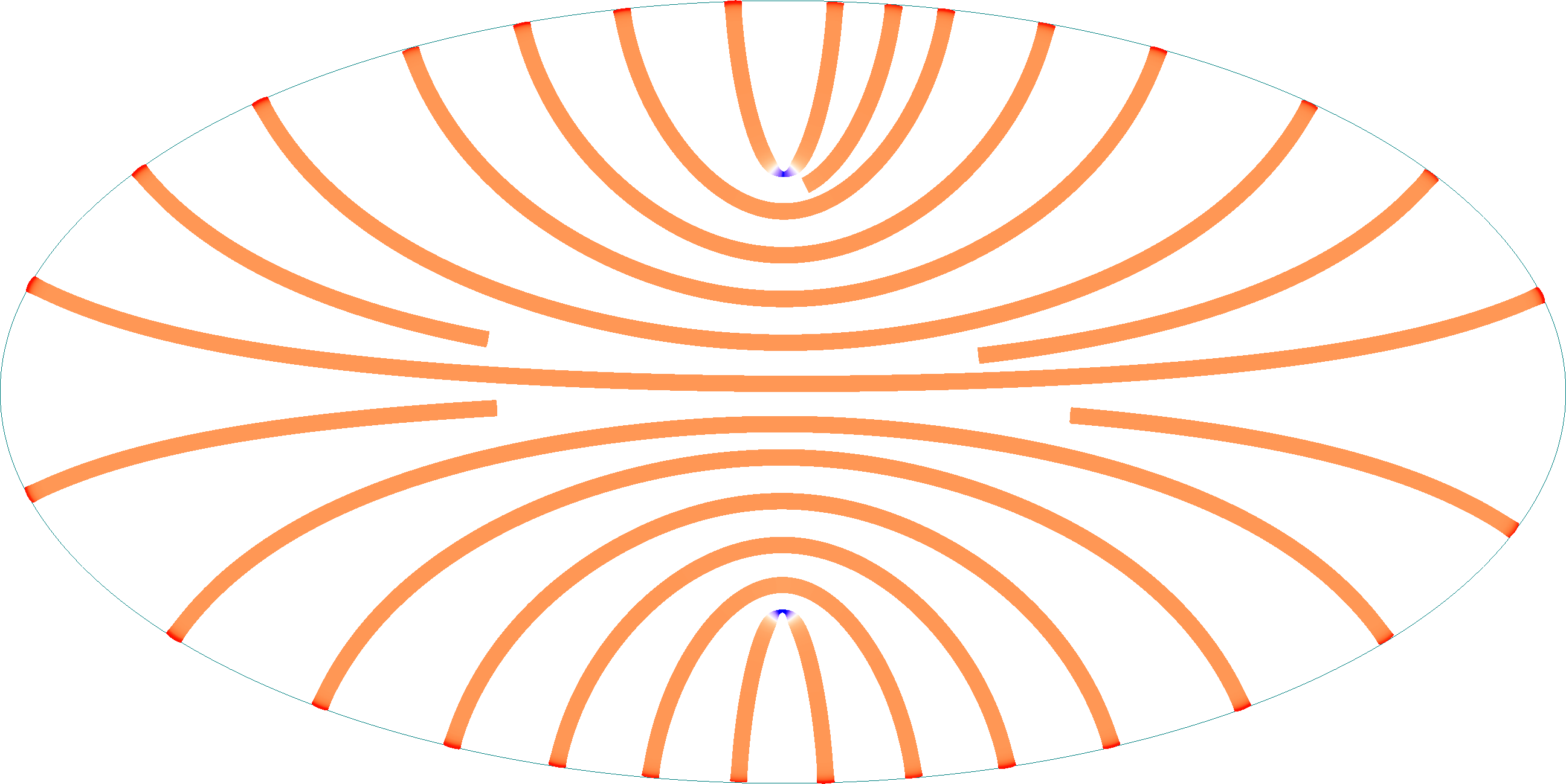}
        \caption{}
    \end{subfigure}
    \begin{subfigure}[b]{0.31\linewidth}        
        \includegraphics[width=\linewidth]{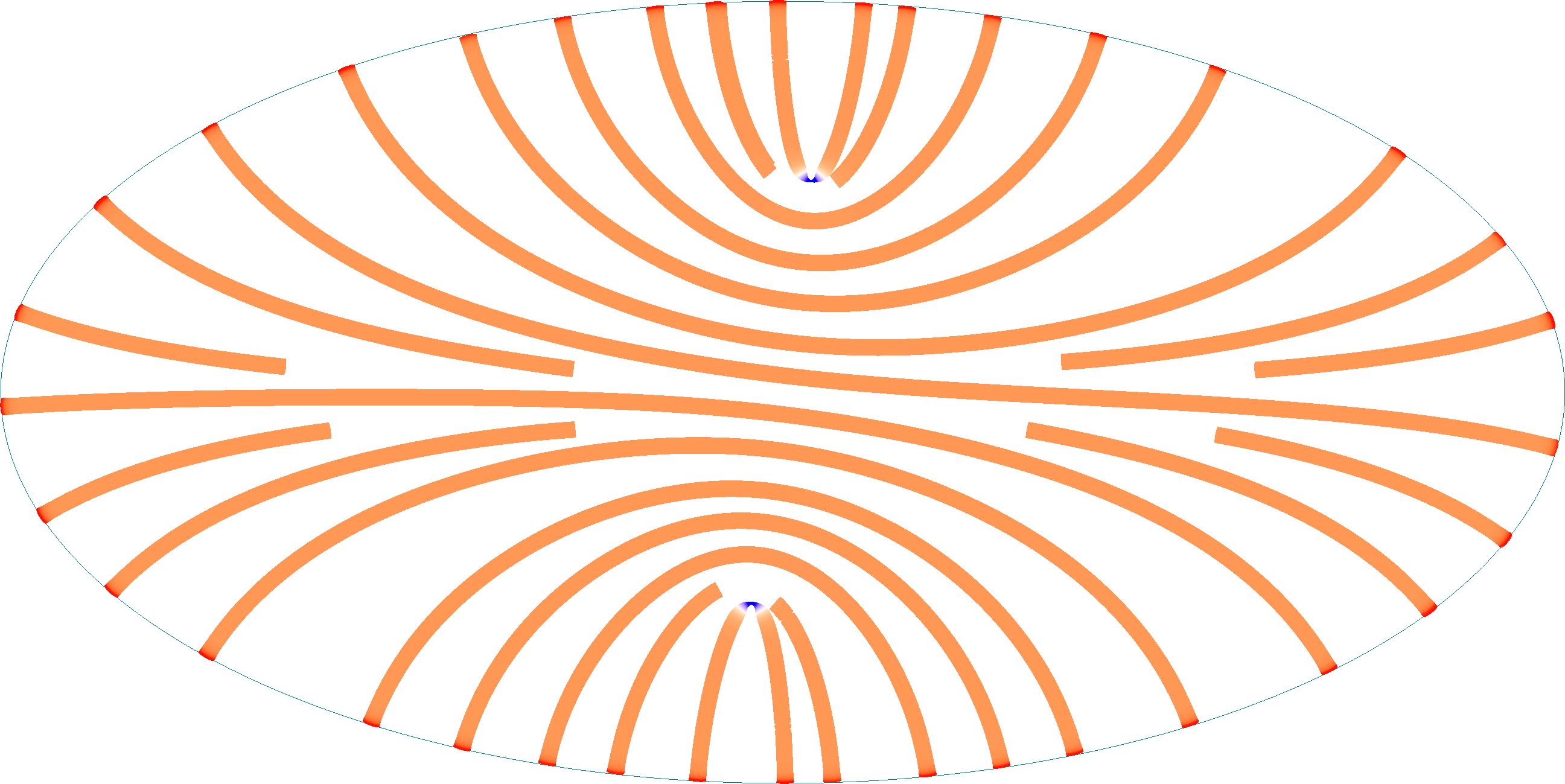}
        \caption{}
    \end{subfigure}
    \begin{subfigure}[b]{0.31\linewidth}    
        \includegraphics[width=\linewidth]{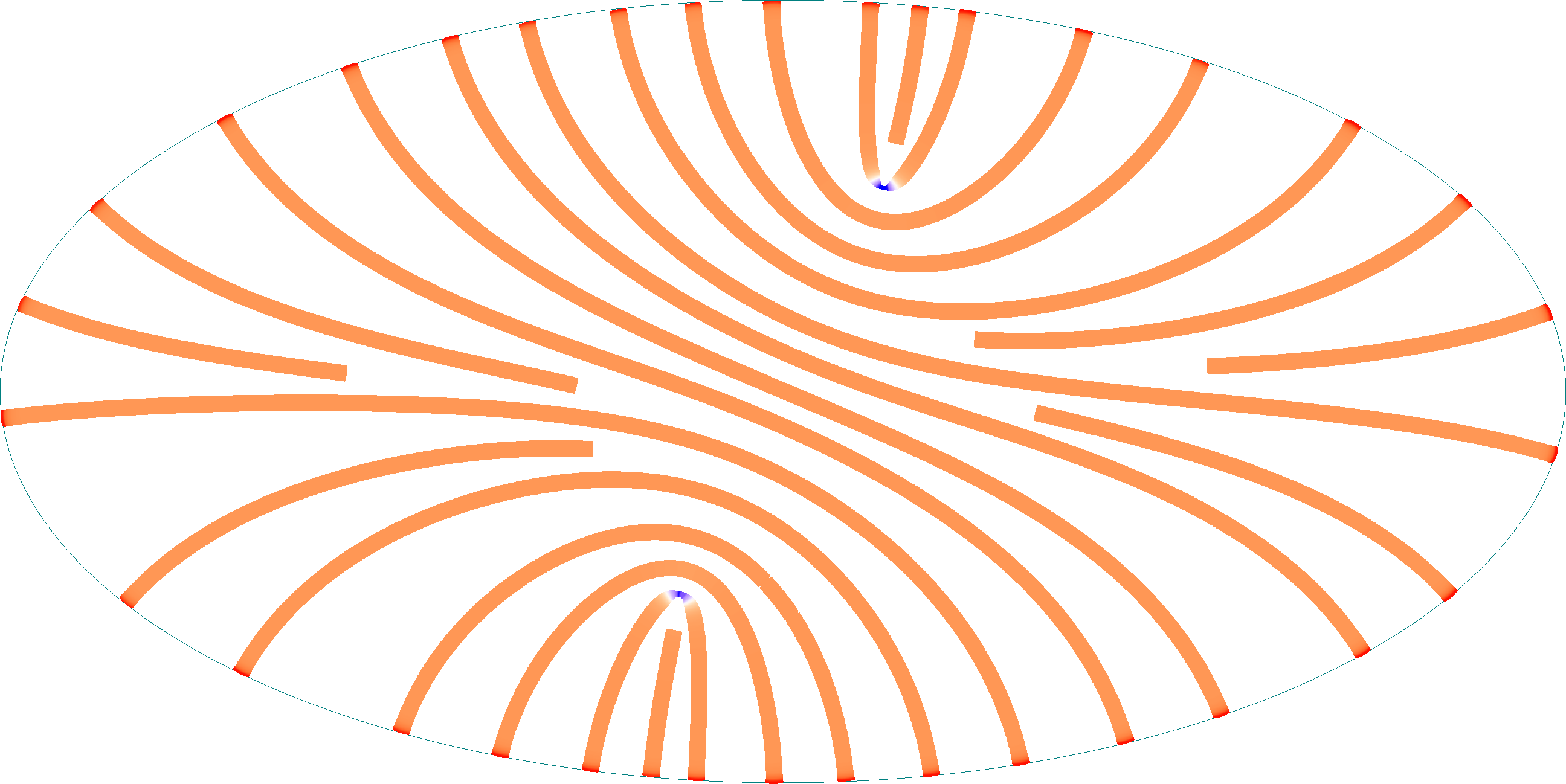}
        \caption{}
    \end{subfigure}\\
    \begin{subfigure}[b]{0.31\linewidth}
        \includegraphics[width=\linewidth]{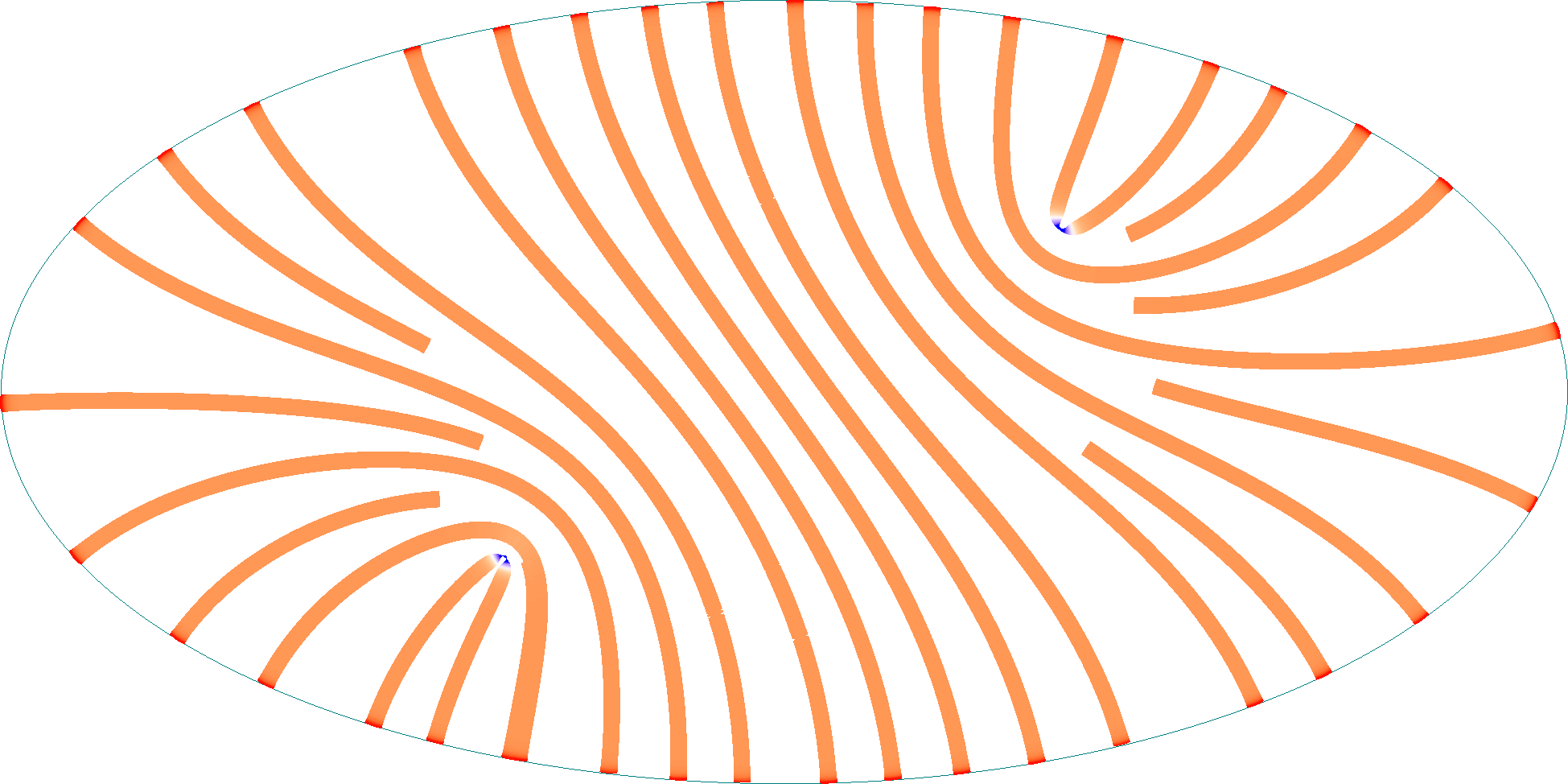}
        \caption{}
    \end{subfigure}
    \begin{subfigure}[b]{0.31\linewidth}        
        \includegraphics[width=\linewidth]{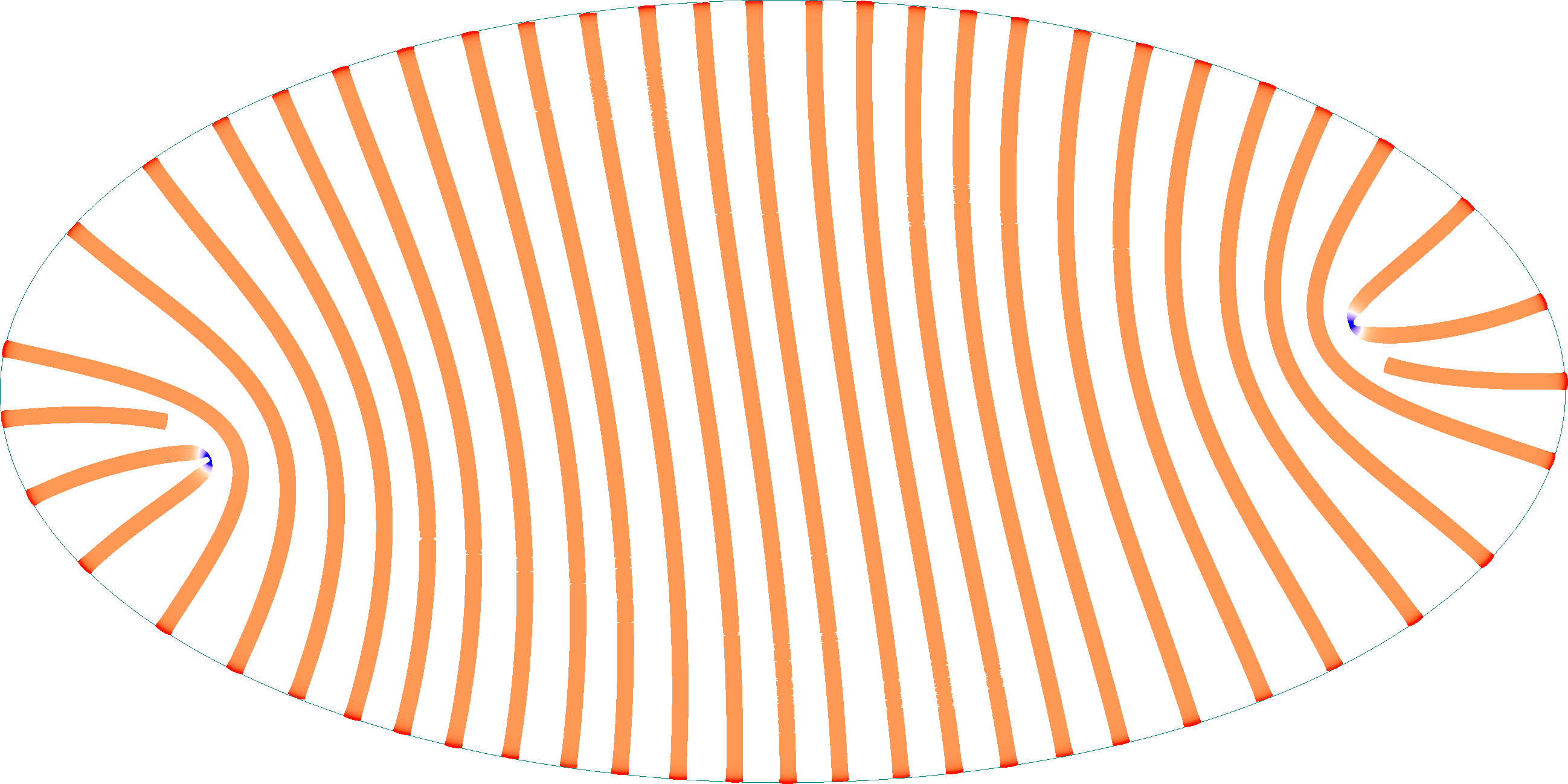}
        \caption{}
    \end{subfigure}
    \begin{subfigure}[b]{0.31\linewidth}    
        \includegraphics[width=\linewidth]{ar2_equilibrium}
        \caption{}
    \end{subfigure}
    \includegraphics[scale=0.07]{scalebar_horiz}
    \caption{(Color online) Hyperstreamline visualizations of the release texture dynamics for the $R=2$ domain, continued from Fig.~\ref{fig:driven_dynamics}i ($t=0$), with  $t = \{\SI{1.79e6}{}$, $\SI{6.75e6}{}$, $\SI{8.55e6}{}$, $\SI{1.03e7}{}$, $\SI{1.26e7}{}$, $\ge \SI{1.60e7}{}$.\label{fig:release_dynamics}}
\end{figure}

Fig. \ref{fig:release_dynamics} shows simulation results of the domain dynamics upon release of the field.
The regimes observed following release of the electric field are found to be similar to that of driven mechanism, but occurring in reverse.
There are two significant differences: timescale and droplet director evolution.
The timescales associated with each regime are an order of magnitude larger compared to their electric field-driven analogues, which is expected due to the different in magnitudes of the surface anchoring and electric field strengths.
The surface anchoring strength governs the magnitude of the restoring forces which, in turn, govern the timescale for nematic orientation dynamics.
They can be compared through their characteristic lengths, where $\frac{\lambda_{s}}{\lambda_{e}} \approx 10$ for the simulation conditions used.
Additionally, during the disclination/bulk rotation regime the droplet director $\bm{n}_{d}$ continuously rotates instead of exhibiting a sharp transition.
This is due to the fact that the droplet scalar order parameter $S_{d}$ does not decrease to zero during growth/recession regime.
Thus a continuous rotation is required in order for the droplet director to return to the equilibrium configuration, orthogonal to the field direction.
Equilibrium textures in all restoration simulations were found to be the same as those from the formation simulations, indicating that under the simulation conditions the electric-field induced deformations of the nematic domain were viscoelastic.

\section{Conclusions}

In this work, a simulation-based study was performed on the formation and electric field switching dynamics of elliptic cylinder nematic domains.
The observed nematic reorientation dynamics were found to have a complex dependence on geometry (aspect ratio), surface anchoring strength, and external field strength.
Both formation and reorientation dynamics were found to be governed by the presence and motion of nematic disclination defects within the domain.
Geometry of the domain, specifically aspect ratio, was found to strongly effect domain texture by providing regions of high curvature to which nematic defects are attracted.
Simulations also predict the presence of a geometry-controlled transition from nematic order enhanced by an external field (low aspect ratio) to nematic order frustrated by an external field (high aspect ratio). 

Equilibrium and dynamic behavior of elliptic nematic domains are found to significantly differ from circular ones, which opens up new possibilities for electro-optical mechanisms for applications of nematic-filled capillaries.
Experimental validation of the presented results is needed, which can be achieved through comparison with future electro-optical dynamics measurements of nematic-filled capillaries as has been done for PDLC films \cite{Drzaic1988}.
Finally, the results presented support the use of simulation-based methods for \emph{rational} design of PDLC optical functional materials tailored to application-specific requirements for nematic texture and switching dynamics.

\begin{acknowledgments}
This work was made possible by the Natural Sciences and Engineering Research Council of Canada (NSERC) and Compute Ontario. 
\end{acknowledgments}

\appendix

\section{Initial Conditions for Heterogeneous Nucleation}\label{sec:app1}

The initial conditions for all simulations assume a boundary layer that is uniaxial and aligned with the surface normal $\bm{k}(\theta)$, where $(r, \theta)$ are the polar coordinates of each surface point.
A linear decay was used for the uniaxial nematic order parameter $S$ (eqn. \ref{eqn:q_tensor}):
\begin{equation}
S_{init}(r, \theta) = S_{b}\left( \frac{r - r_{s}(\theta) + \lambda_{n}}{\lambda_{n}} \right) \qquad  \textrm{for } \qquad r_{s}(\theta) - \lambda_{n} < r < r_{s}(\theta )     
\end{equation}
\begin{equation}
S=0   \qquad \textrm{for} \qquad  0 \le r \le  r_{s}(\theta) - \lambda_{n}     
\end{equation}
where $S_{b}$ is the bulk order parameter value at the simulation temperature, $\lambda_{n}$ is the nematic coherence length (eqn. \ref{eqn:coherence_length}), and $r_{s}(\theta)$ is the radial coordinate of the ellipse surface.
The expression for $r_{s}(\theta)$ is:
\begin{equation}
r_{s}(\theta)= \frac{a}{\sqrt{b^{2}\cos^{2}\theta + a^{2}\sin^{2}\theta }}    
\end{equation}
where $a/b$ is the length of the major/minor ellipse axis.
The nematic director field (eqn. \ref{eqn:q_tensor}) is assumed to be well-aligned with the ellipse surface normal:
\begin{equation}
    \bm{n} = a\cos(\theta) \bm{e}_{1} + b\sin(\theta) \bm{e}_{2}
\end{equation}
resulting in the initial Q-tensor field: 
\begin{equation}\label{eqn:3:initial_conditions} 
\bm{Q}_{init} =S_{b} \left(\bm{nn} - \frac{1}{3} \bm{\delta}\right)
\end{equation}

\bibliography{self_assembly,pdlcs,computational}

\end{document}